%% file: V3_arxiv.tex
\documentclass[12pt, one column]{IEEEtran}

\usepackage[numbers]{natbib}
\usepackage[utf8]{inputenc}
\usepackage{tikz}
\usepackage{pgfplots}
\usepackage{tikz-3dplot}
\usetikzlibrary{calc,patterns,angles,quotes}
\tdplotsetmaincoords{60}{115}
\pgfplotsset{compat=newest}

\usepackage{lineno}
\usepackage{tabularx}
\usepackage{lipsum}
\usepackage{mathtools}
\usepackage{amsmath,amssymb,amsfonts}

\usepackage{graphicx}
\usepackage{subcaption}
\usepackage{textcomp}
\usepackage{soul}
\usepackage{color}
\usepackage{caption}
\usepackage{floatrow}
\usepackage{wrapfig}
\usepackage{comment}
\usepackage{pifont}
\usepackage{float}

\usepackage{xcolor}
\usepackage{enumitem}

\usepackage{hyperref}
\modulolinenumbers[5]
\usepackage{pdflscape}
\usepackage{natbib} 
 \DeclarePairedDelimiter\ceil{\lceil}{\rceil}
 
\DeclarePairedDelimiter\abs{\lvert}{\rvert}%
\usepackage{algorithm}
\usepackage{algpseudocode}

\begin{document}

\title{Fastening the Initial Access in 5G NR Sidelink for 6G V2X Networks}
\author{Marouan Mizmizi, Francesco Linsalata, Mattia Brambilla, Filippo Morandi, Kai Dong, Maurizio Magarini, Monica Nicoli, Majid Nasiri Khormuji, Peng Wang, Renaud Alexandre Pitaval, and Umberto Spagnolini
\thanks{Marouan Mizmizi, Francesco Linsalata, Filippo Morandi, Kai Dong, and Maurizio Magarini are with Dipartimento di Elettronica, Informazione e Bioingegneria, Politecnico di Milano.\\
Mattia Brambilla and Monica Nicoli are with Dipartimento di Ingegneria Gestionale, Politecnico di Milano.\\
Majid Nasiri Khormuji, Peng Wang, and Renaud Alexandre Pitaval are with Huawei Technologies Sweden AB.\\
Umberto Spagnolini is with Dipartimento di Elettronica, Informazione e Bioingegneria, Politecnico di Milano, and he is Huawei Industry Chair.}}

\maketitle

\begin{abstract}
The ever-increasing demand for intelligent, automated, and connected mobility solutions pushes for the development of an innovative sixth Generation (6G) of cellular networks. A radical transformation on the physical layer of vehicular communications is planned, with a paradigm shift towards beam-based millimeter Waves or sub-Terahertz communications, which require precise beam pointing for guaranteeing the communication link, especially in high mobility. A key design aspect is a fast and proactive Initial Access (IA) algorithm to select the optimal beam to be used. In this work, we investigate alternative IA techniques to fasten the current fifth-generation (5G) standard, targeting an efficient 6G design. First, we discuss cooperative position-based schemes that rely on the position information. Then, motivated by the intuition of a non-uniform distribution of the communication directions due to road topology constraints, we design two Probabilistic Codebook (PCB) techniques of prioritized beams. In the first one, the PCBs are built leveraging past collected traffic information, while in the second one, we use the Hough Transform over the digital map to extract dominant road directions. We also show that the information coming from the angular probability distribution allows designing non-uniform codebook quantization, reducing the degradation of the performances compared to uniform one. Numerical simulation on realistic scenarios shows that PCBs-based beam selection outperforms the 5G standard in terms of the number of IA trials, with a performance comparable to position-based methods, without requiring the signaling of sensitive information.
\end{abstract}


\section{Introduction}
Connected mobility is a flagship element of  smart cities and a mandatory step in the evolution towards automated driving, to guarantee traffic safety, efficiency, user comfort and environmental sustainability~\cite{SAE_J3016_201806,BobKouManEicXu:J18,CamMolSco:J15,QuWanYan:J10}. 
Moreover, sales volume and market share of connected vehicles represent a radical breakthrough for telecommunication operators, unlocking new business models and strategies \cite{MahShoMccPapEluOlu:J21}.
In this framework, vehicular communications represent the key technology to enable information sharing and thereby cooperation among road users, to augment the ego-vehicle sensing and control capabilities.
The heterogeneity of envisioned mobility applications, spanning from driving-assistance to on-board entertainment~\cite{ZhaCheLyuCheShiShe:J18}, call for a vehicular network that is extremely reliable, fast and resilient.

The European Telecommunications Standards Institute (ETSI) has identified connected services for active road safety and cooperative traffic efficiency, to be enabled by Vehicle-to-Everything (V2X) communications, either Vehicle-to-Vehicle (V2V) or Vehicle-to-Infrastructure (V2I), since 2009~\cite{ETSI_BasicV2Xservices}. Therein, the addressed use cases belong to the low levels of automation (levels 1 and 2, according to~\cite{SAE_J3016_201806}), i.e., driver assistance and partial automation features.  
To enable higher automated driving features, enhanced V2X (eV2X) scenarios have been defined in~\cite{3GPP_5Greq2020}, with stringent requirements that currently available V2X communication technologies, i.e., IEEE 802.11p~\cite{802_11p} and Long Term Evolution (LTE) V2X~\cite{LTEV2X}, cannot meet~\cite{StoDua:J20,SRIVASTAVA:evc20,SinNanNan:EVC19}. 
In this regard, the recent roll-out of fifth Generation (5G) technology has pushed for an advanced telecommunication standard capable of handling diverse applications and related requirements. In its design phase the automotive industry has been considered as a key vertical sector and it is foreseen that with a pervasive 5G deployment it can start the (r)evolution of connected mobility. However, it is more likely that fully automated and connected vehicles will be initiated by sixth Generation (6G), where extreme data rates ($\sim\! 1 \!$ Tbps) and zero-latency ($<\!1$ ms) are foreseen~\cite{SaaBenChe:J20}, to be enabled by shifting towards millimeter Waves (mmWaves) or sub-THz frequencies, even for V2V communications. 
The pervasive deployment of 6G V2X devices will guarantee a seamless connectivity among the vehicles for real-time exchange of sensor data streaming (camera, lidar, radar, etc.) and synchronization of the driving trajectories, improving road safety and traffic efficiency. Furthermore, in addition to cooperative driving, the 6G vehicular platform will also be the basis to deliver a plethora of ad-hoc user-centred services, for an augmented mobility experience as a whole.

The envisioned revolutions requires a radical transformation of the current physical layer for V2X communications and advanced processing techniques for precise spatial beamforming. Narrow-beam communications are deemed as candidate solution for high spectral and energy efficiency requirements. 
This is the research area targeted in this paper, where we investigate how beam pointing is conceived in the current standard, and we assess potential enhancements based on side statistical information extracted from maps. Considering that 6G research is still it its early days, in this paper we start from detailing the current available 5G standard and we then discuss potential improvements on the beam selection mechanism for the  Initial Access (IA) in V2V towards highly-directional beam control for 6G design.

\subsection{Related Works}
The IA problem represents a hot research topic, and the multitude of literature works confirm its relevance for practical systems, as surveyed in  \cite{WonCho:J20} where an historical analysis on UE discovery in mobile communication systems is given, from 3G to 5G. With the 6G trend of increasing the transmission frequency (mmWave and sub-THz), and reducing the beamwidth, the problem of beam selection becomes of paramount importance, and efficient solutions must be designed. Alternative approaches to 5G NR standard have been recently proposed in many research works, for a generic User Equipment (UE) connection to a base station~\cite{SimLimParDaiCha:J20,SolParTomZor:J19,OmrShaAliAln:J19,MezZhaPolForDutRanZor:J18,LiuAuMaaLuoBalTonChaLor:J18,RagCezSubSamKoy:J16,ShoFisFodPopZor:J15,BarHosRanLiuKorPanRap:J15} as well as specifically intended to V2X applications, both in V2I~\cite{RASHEED:evc21,Nordio:J2020,RasHuBal:J20,BrambillaICC2020,MunsInfocom2019,Hanzo:J2019,WanKlaRibSooHea:J19,KlaGonHea:L19,SimKloAsaKleHol:J18,Va:J2018,Mavromatis:C2017,Hollick:C2017,Va:C2016} and V2V~\cite{Brambilla2020sensors,Perfecto:J2017,Feng:C2019} contexts.

In the following, we consider the literature on V2X communications for coherence with the scopes of this paper, first discussing the solutions for V2I, followed by the more challenging V2V case.
To speed up the IA in V2I communications, in \cite{RASHEED:evc21} a combination of Software Defined Networking (SDN)-controlled and Cognitive Radio (CR)-enabled V2X routing is proposed, where a multi-type2 fuzzy inference system  (M-T2FIS) is used for the optimal beam selection enabling a switching between mmWave and THz technologies.
A-priori road topology and traffic signal information is used in~\cite{Nordio:J2020}, with the goal of maximizing the aggregate throughput of a group of served vehicles. The idea of using an information that is available beforehand (i.e., it is an input to instantaneous beam selection) is also explored in ~\cite{Hanzo:J2019,Va:J2018} where they query a  fingerprint database containing a prior information of candidate pointing directions, which is continuously updated by learning. A similar database-based construction by learning and query is also proposed in ~\cite{BrambillaICC2020}, where the beam selection relies on eigen-beamformers that exploit  sparsity properties of the mmWave V2I channel, and in \cite{SimKloAsaKleHol:J18} where a contextual online learning algorithm addressing the problem of beam selection with environment awareness. The selection mechanism of a mmWave beam can also be assisted by another technology such as radar ~\cite{MunsInfocom2019} or lidar~\cite{KlaGonHea:L19}.
Another approach is to use vehicle position information as sufficient condition of beam alignment~\cite{RasHuBal:J20,Va:C2016}, or as a part of a richer set of environmental features~\cite{WanKlaRibSooHea:J19}, possibly including motion prediction and V2I distance estimation for beam adaptation~\cite{Mavromatis:C2017}.

In the V2V case, the beam selection problem has been addressed by the integration of on-board position and inertial sensors in~\cite{Brambilla2020sensors} for unknown position and orientation of both transmit (Tx) and receive (Rx) vehicles, or by using channel and queue state information to optimize both transmission and reception beamwidth~\cite{Perfecto:J2017}. It has also been formulated as optimization problem to maximize the sum of the average transmission throughput in the nearby regions of the Rx vehicle under Tx-only uncertainty assumption (but without orientation uncertainty) \cite{FenHeGuaHuaXuChe:J21}.

\subsection{Paper Contributions} 
This paper focuses on the analysis of IA techniques in V2V communications. The 5G NR solution is considered as reference algorithm, and enhancements to fasten the IA phase are proposed with a look to 6G era and its low latency requirements. Part of the concepts we present in this work have been recently published in \cite{LinsICC-c21}. With respect to that work and the current literature, the new contributions of this paper can be summarized as follows.
\begin{itemize} [wide]
\item [\textit{i})] Most of mentioned works lack of realistic vehicular scenarios with FR2 mmWaves channel modeling. The scenario is here simulated by using OpenStreetMap~\cite{osm} and Simulation of Urban MObility (SUMO)~\cite{SUMO2018} software, whose combination allows an accurate modeling of vehicle traffic over real road networks. Furthermore, the SUMO output is processed by Geometry-based, Efficient propagation Model for Vehicle-to-Vehicle (V2V) communication (GEMV$^\mathbf{2}\,$)~\cite{GEMV2} that has been adapted to account for the computation of mmWave  channel according to the 3rd Generation Partnership Project (3GPP) guidelines in~\cite{14rel}. 
\item [\textit{ii})] The review of different beam-sweeping techniques suggests to use the position of the vehicles to fasten the IA phase. We show how an incorrect sidelink positioning can easily decrease the V2V link Signal-to-Noise Ratio (SNR). Two approaches  are proposed to take into account for the localization error.  The first approach is based on left-right jump around the signaled position, the second one implements an iterative procedure that aims to maximize the received power.
\item [\textit{iii})] A statistical analysis of Angles of Arrival (AoA) and Departure (AoD) highlights that V2V communications exhibit preferential angular directions related to road topology and traffic conditions. This motivates the design of a beam selection method based on a Probabilistic Codebook (PCB) approach for selecting the optimal beam direction, which emerges as a valid alternative to 5G NR exhaustive search and suggests possible directions for 6G V2V standards. We design two  methods to determine the PCB in case of urban and highway scenarios. One is based on training phase over repetitive road-dependent traffic, the other relies on the so-called Hough Transform to extract the dominant road directions from digital maps of the driving environment. We analyze the benefits and limits of both of them.
\item [\textit{iv})] Simulation results over realistic traffic environments are proposed to validate the developed methods. In particular, numerical results show improvements using a PCBs in terms of average minimum number of trials with respect to the 5G NR approach, achieving the upper-bound performances of position-assisted (e.g., through Global Positioning System - GPS) schemes in some case. Furthermore, performance loss due to angle quantization for codebook design appear as inevitable. Here, we use different quantization approaches, i.e., uniform and irregular quantization, and compare with the optimal one using Singular Value Decomposition (SVD) of the channel matrix to investigate the 
\end{itemize}

\subsection{Paper Organization} 
The remainder of the paper is structured as follows. The 5G NR standard and its IA procedure are discussed in Sec.~II. 
The system model and simulations methodologies, which groups both the V2V communications setting and the computer simulations design/implementation/parameters, is presented in Sec.~III. Sec. IV outlines the beam selection schemes investigated in this paper, detailing the implementation and discussing pros and cons of both position-assisted schemes and PCBs-based ones. The numerical results are reported in Sec.~V. Lastly, Sec. VI summarizes and concludes the work.

\section{Overview of 5G NR Standard}
\label{sec:IAin5GNR}
In this section, we provide an overview on the current 5G New Radio (NR) standard. We start by introducing the latest version of the standard (Release 16) and discussing its key features in subsection \ref{subs:5G V2X},  with main target to V2V communications. Then, in subsection \ref{subs:IAin5GNR}, we analyze the IA problem, describing how it is addressed in 5G NR, with  focus on technical aspects of the physical layer.

\subsection{Key Features of 5G NR Sidelink}\label{subs:5G V2X}

The 3GPP Release 16, known as 5G NR, has been recently specified for both the uplink~/~downlink (i.e., V2I) and sidelink (i.e., V2V) communication modes~\cite{GarMolBobGozColSahKou:J21,Cheng:J20,TS_38213}. With this release, the Uu (uplink/downlink) and PC5 (sidelink) interfaces of LTE  are replaced by a brand new version entirely based on 5G NR air interface, thus setting the first milestone for future V2X standards~\cite{ParBlaBlaDahFodGraStaSta:J20}. Indeed, 5G NR can operate in strict interoperability with LTE networks (non-standalone mode), i.e., using the LTE radio access network as an overlay,  or independently (standalone mode) \cite{6GFlagship,GioPolRoyCasZor:J19}.

Among the 5G NR novelties, the most significant features of 5G-V2X reside in the introduction of (\textit{i}) unicast and groupcast transmissions, which expand the broadcast ones; (\textit{ii}) a dedicated physical sidelink feedback channel (PSFCH), which complements the physical sidelink broadcast channel (PSBCH), the physical sidelink control channel (PSCCH), and the physical sidelink shared channel (PSSCH); (\textit{iii}) a flexible numerology, which allow for transmissions at different frequencies, either at sub-$6$ GHz, i.e., Frequency Range 1 (FR1), or mmWaves, i.e., FR2~\cite{GanLohMalKunKuc:J20,GarGonMarMon:J20,ZeaJavHam:J20}.

The 5G NR at FR2 embodies the latest frontier of cellular technology, and its feasibility is corroborated by practical demonstrations~\cite{NohKimChoLeeChuKim:J21,LiYuFukTraSak:J21,Kim:J19,Hollick:C2017,Rappaport:J13,1Xinyu2020}. The experimentation, jointly with theoretical studies on electromagnetic propagation, emphasize the need of Multiple-Input Multiple-Output (MIMO) antenna arrays as a mandatory hardware technology to confine spatial radiation through beamforming and compensate for high path loss.

A main distinctive feature of 5G NR design is to accommodate for directional communication by introducing beam selection mechanism, in which an optimal pair of transmit and receive beams (among many candidates) is determined \cite{LieShiHuaSuHsuWei:J17}. The choice of the beam is carried out at the first connection, i.e., IA, and whenever a link failure is detected, while if the V2X communication link is already established, a beam tracking mode takes place~\cite{Huang:J2020,GioPolRoyCasZor:Sur19}.
The IA procedure in 5G NR standard is implemented by periodic transmission of synchronization signals, which are selectively transmitted over different beam directions through spatial beamforming \cite{TS_38213}. However, this procedure is widely acknowledged to be inefficient for future releases of the standards, especially if very narrow beams are used  and, most notably, if it has to operate with high terminal mobility ($500$ km/h in 5G \cite{3GPP_5Greq2020}, and $1000$ km/h in 6G \cite{JiaHanHabScho:J21}).

\subsection{Initial Access in 5G NR Sidelink}\label{subs:IAin5GNR}

\begin{figure}[b!]
    \centering
    \includegraphics[width=0.7\columnwidth]{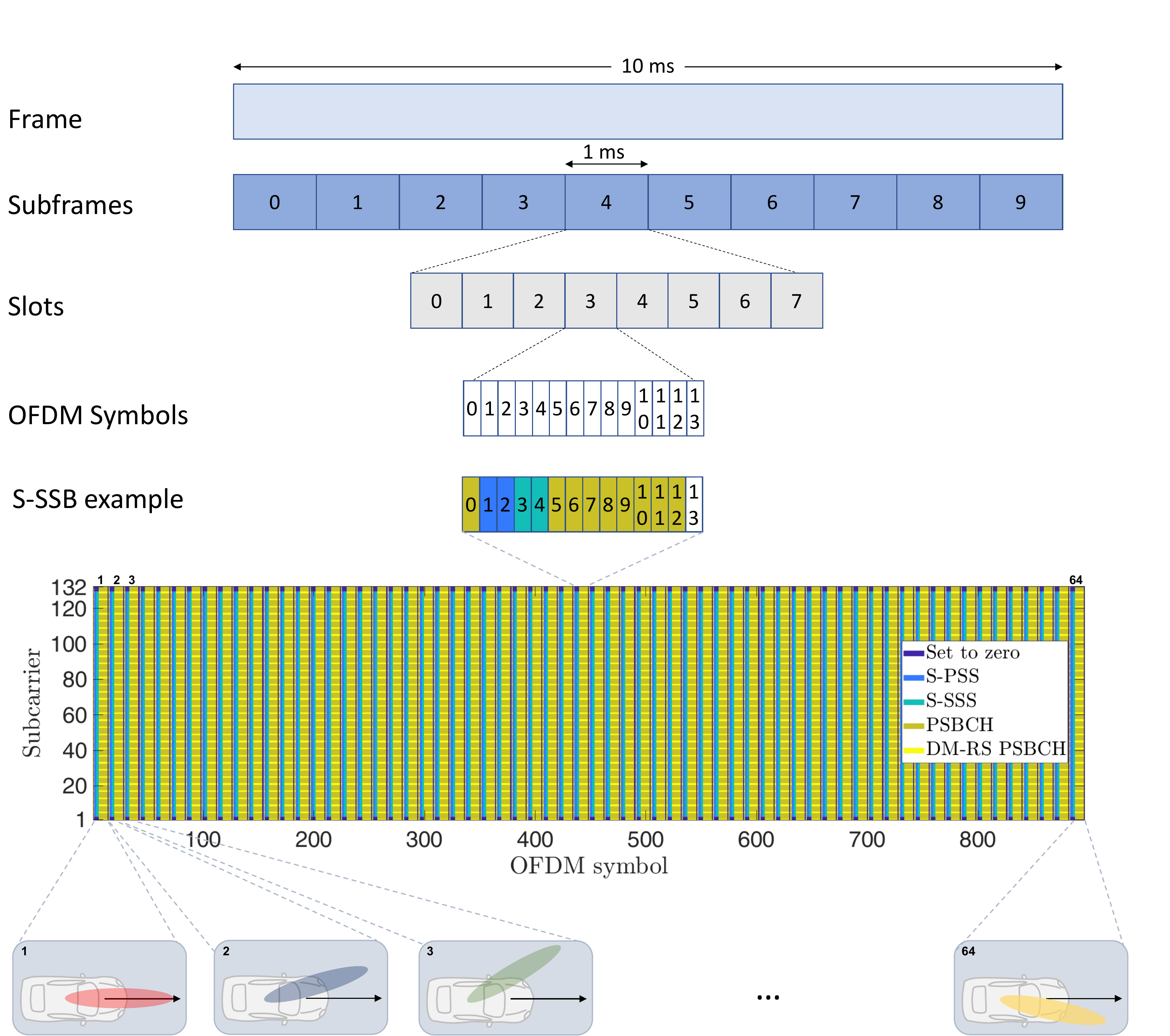}
    \caption{Example of a 5G NR frame structure with division in subframes, slots, and OFDM symbols. The FR2 numerology $\mu=3$ is considered, where eight slots compose a subframe. An example of S-SSB is also reported, showing the match with a spatial beam and how it frames into an S-SS burst (an aggregation of $N_s=64$ S-SSB here).}
    \label{fig:5Gstandard}
\end{figure}

In this subsection, we detail how the IA procedure is carried out in the 5G NR Rel. 16~\cite{TS_38213}, which represents the acknowledged reference method and benchmark. According to the standard, as for uplink/downlink, in sidelink transmission the communication is organized in frames of $10$ ms, each of them composed of  $10$ subframes of $1 \!$ ms (see Fig.~\ref{fig:5Gstandard}). The 5G NR standard allows high flexibility in spectrum sharing by enabling different numerologies ($\mu=0,1,2,3,4$), which provide a scalable Sub-Carrier Spacing (SCS). In the current release~\cite{TS_38213}, FR2 sidelink can only support SCS of $120 \!$ KHz. Therefore, a subframe can accommodate $8$ slots, each with $14$ Orthogonal Frequency-Division Multiplexing (OFDM) symbols with Normal Cyclic Prefix (NCP).

The IA procedure in 5G NR is detailed for a UE connecting to a gNB, i.e., for uplink/downlink transmission. For sidelink, however, a similar procedure takes place. Thus, we will use the following terminology for clarity: we use the notions of Vehicle UE (V-UE) and Vehicle Tx (V-Tx) to indicate the receiver and the transmitter, respectively.

Sidelink beam selection occurs at the physical layer by the periodic transmission/reception of Sidelink Synchronization Signals (S-SSs) at dedicated frequency locations. 
S-SSs are sent from a V-Tx and they convey synchronization information in the form of sidelink primary synchronization signal (S-PSS) and sidelink secondary synchronization signal (S-SSS), which, together with the PSBCH, constitutes a Sidelink Synchronization Signal Block (S-SSB). 
In Rel. 16, an S-SSB occupies one entire slot (an example is given in Fig.~\ref{fig:5Gstandard}). Its periodicity is set to $16$ frames, i.e., every $160 \!$ ms, and, in the frequency domain, it occupies  $11$  resource blocks of $12$ subcarriers each (i.e., 132 subcarriers overall). Note that S-SSB transmission is outside the resource pool (the subset of available resources in time/frequency domains for sidelink transmission).
Each S-SSB is beamformed to a specific spatial direction, as illustrated in Fig. \ref{fig:5Gstandard}, and for numerology $\mu=3$, the number  of S-SSB transmissions $N_s$ can be of $1, 2, 4, 8, 16, 32$ or $64$, each with a different beam.
The aggregation of S-SSB is referred as S-SS burst, and for sidelink its duration, when $N_s=64$, cannot be less than $8$ ms~\cite{R1_2001553}, that corresponds to the case in which the interval between neighboring S-SSBs equals 1 slot and the offset of the first block with respect to the beginning of the period is 0.
The spatial distribution of S-SSB allows a V-Tx to scan all the space domain, guaranteeing a $360^\circ$ coverage.
The V-UE searches for the S-SSB and by decoding them it is able to synchronize and identify the Tx and Rx beams for communicating with the V-Tx. It is out of the scopes of this work to provide details on the decoding part, for which the reader can refer to \cite{GarMolBobGozColSahKou:J21,Cheng:J20}.

From the above discussion, the IA phase reduces to the detection (at V-UE side) of S-SSB,  which are beamformed by a V-Tx. The 5G IA procedure has the advantage of not requiring assistance information (i.e., information from external hardware or software, such as position information of vehicles). However, besides being resources-hungering, it limits the available scanning to only $64$ beams, meaning that it can raise problems if the 6G trend of moving towards extremely high frequency communications, or sub-terahertz, and requiring highly-narrow beams will happen~\cite{JiaHanHabScho:J21,RappaportTHz100GHz:J19}. Moreover, the periodicity of $160$ ms can represent a drawback for high mobility systems, as the variations in the environment can be significant (a vehicle at $130$ km/h travels for $5.7$ m in $160$ ms) and it does not guarantee low-latency communications for safety-critical applications, as required for enhanced V2X services \cite{3GPP_5Greq2020}.

To overcome these limitations, in Sec. \ref{sec:BSS} we discuss how position-assisted cooperative solutions can represent a valid alternative (at the expenses of letting vehicles reciprocally acquire such information), and we propose a probabilistic approach  that extracts statistical knowledge from  nearly-repetitive mobility patterns to construct a codebook of prioritized beam-pointing directions, without requiring assistance information nor the sharing of sensitive data. Before entering into the details of beam selection algorithm, in the next section we define our system model to be used in methodological evaluation.

\section{Simulation Methodology}

\begin{figure}[!t]
    \centering
    \includegraphics[width=0.7\columnwidth]{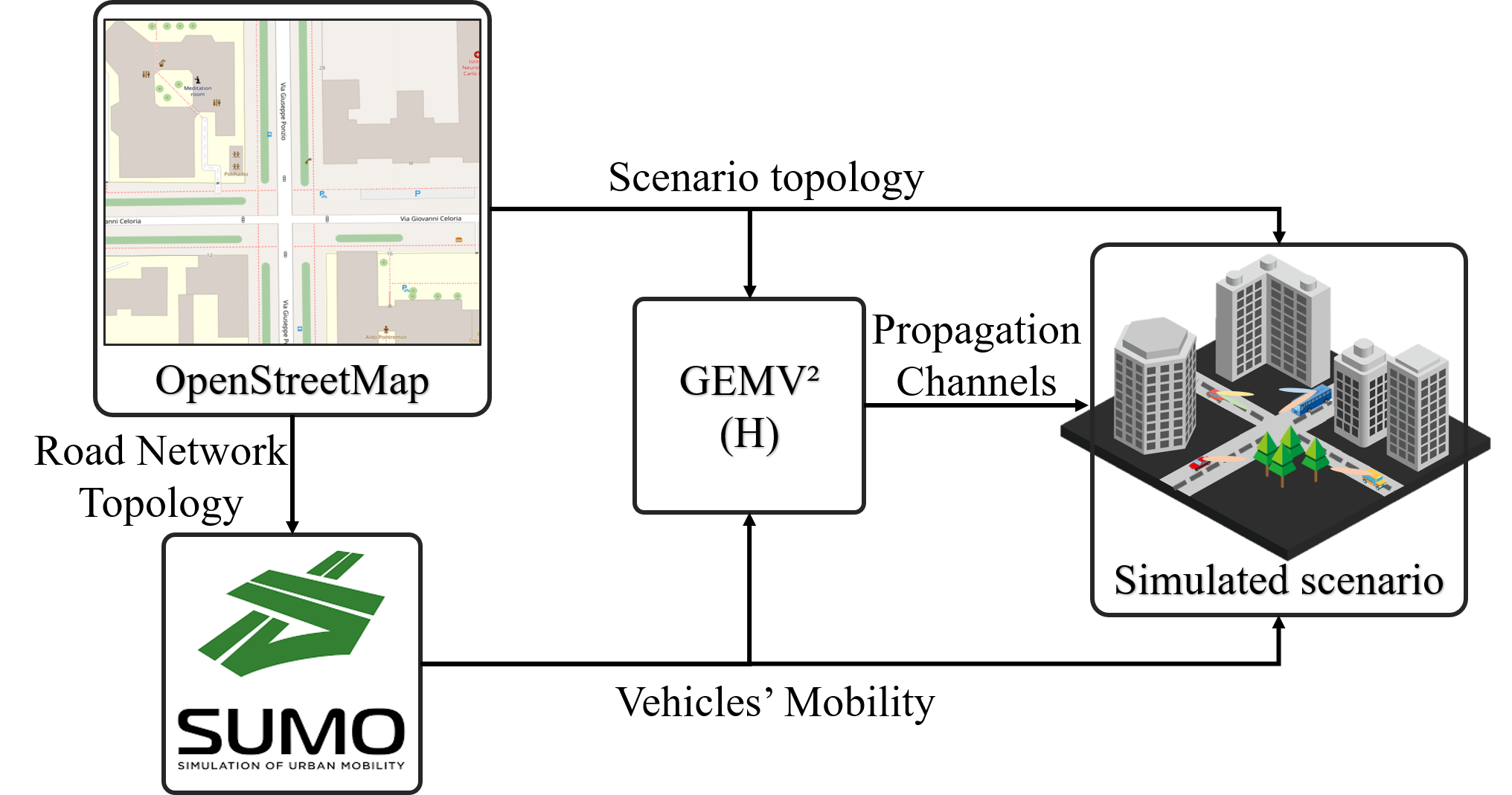}
    \caption{Overall simulation system for evaluation methodologies: the scenario layouts are taken from OpenStreetMap~\cite{osm}, the mobility is modelled by SuMO~\cite{SUMO2018}, GEMV2~\cite{GEMV2} deals with the computation of the mmWave the V2V channel matrix.}
    \label{fig:sysModel}
\end{figure}

The block diagram for the proposed simulation methodologies and settings is depicted in Fig.~\ref{fig:sysModel}, which is intended to model mmWave V2V communications in lifelike mobility scenarios. The objective is to evaluate the beam selection methods in a complex vehicular environment with realistic mmWave propagation. The road network, buildings, and static objects topologies are obtained from realistic digital maps, i.e., OpenStreetMap~\cite{osm}, while SUMO software~\cite{SUMO2018} is adopted to simulate vehicular mobility. The sparse nature of the mmWave channel is captured by employing GEMV$^2$~\cite{GEMV2}, which is a geometry-based channel model. The mmWave propagation parameters are defined based on the 3GPP requirements in~\cite{14rel}.

\subsection{Urban and Suburban Scenarios}

\begin{figure*}[!t] 
\centering
\subfloat[Intersection \label{fig: intersection} \label{fig:intersection}]{\includegraphics[width=0.32\columnwidth]{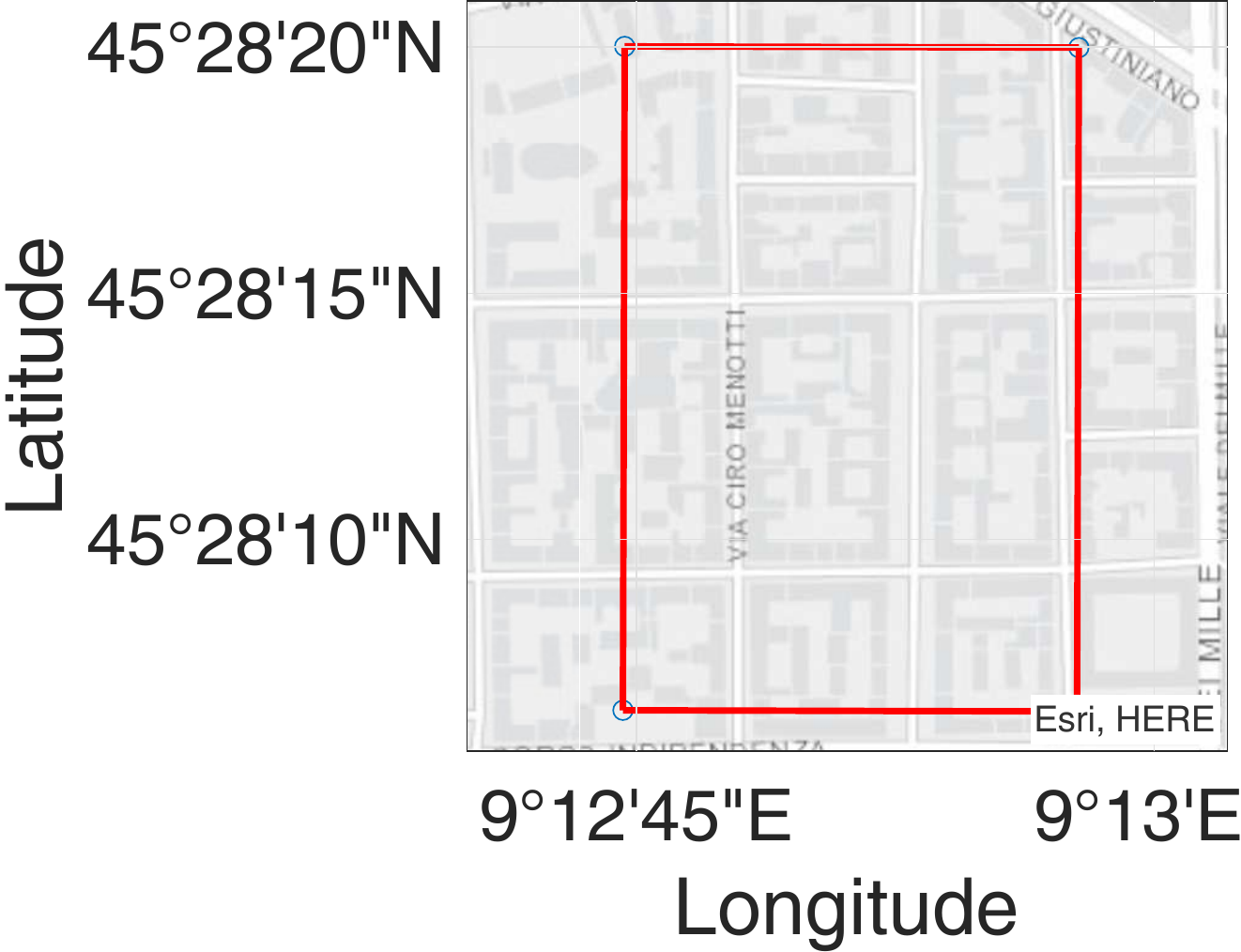}}
\subfloat[Roundabout \label{fig:roundabout}]{ \includegraphics[width=0.32\columnwidth ]{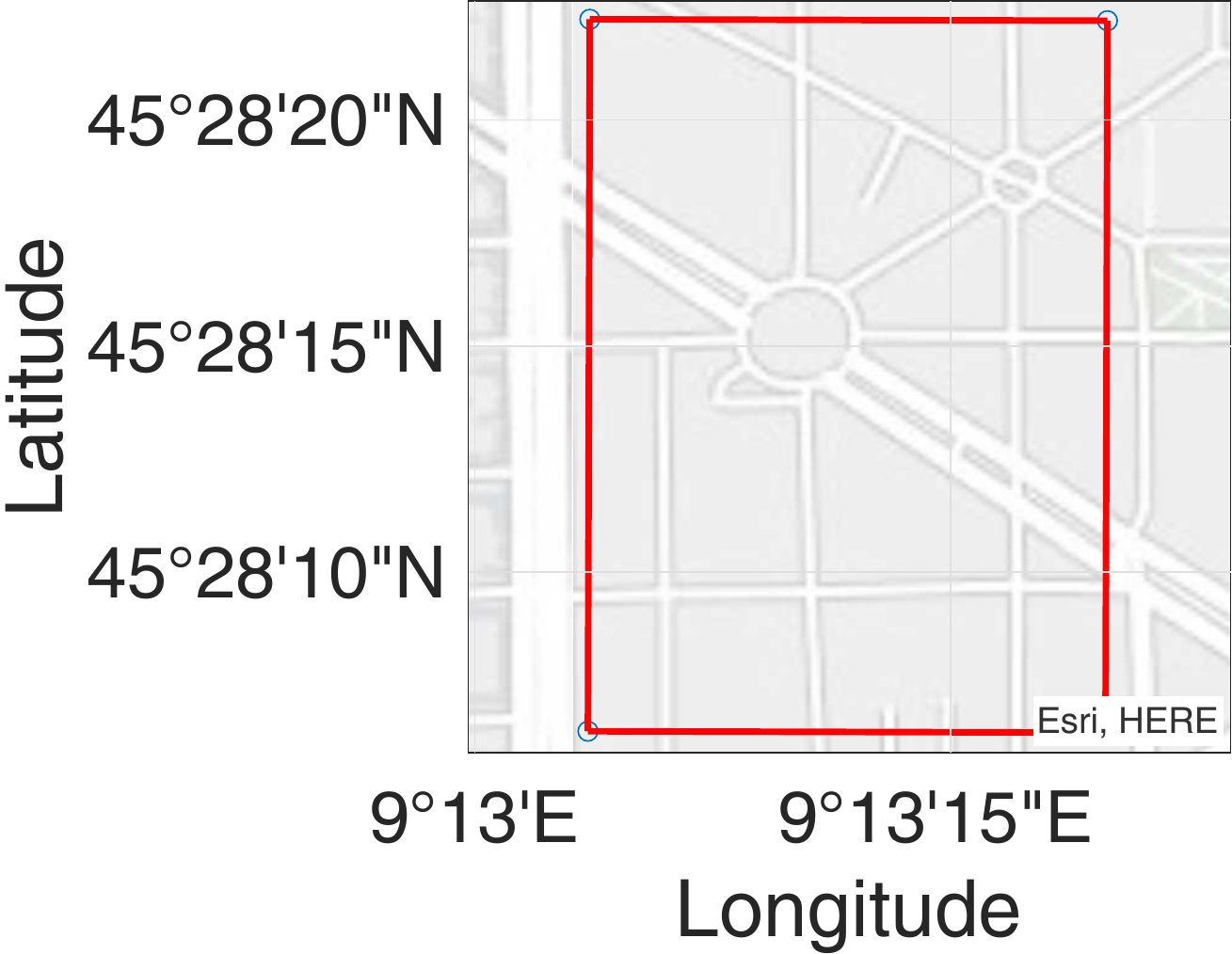}}
\subfloat[Highway \label{fig: highway}]{ \includegraphics[width=0.32\columnwidth]{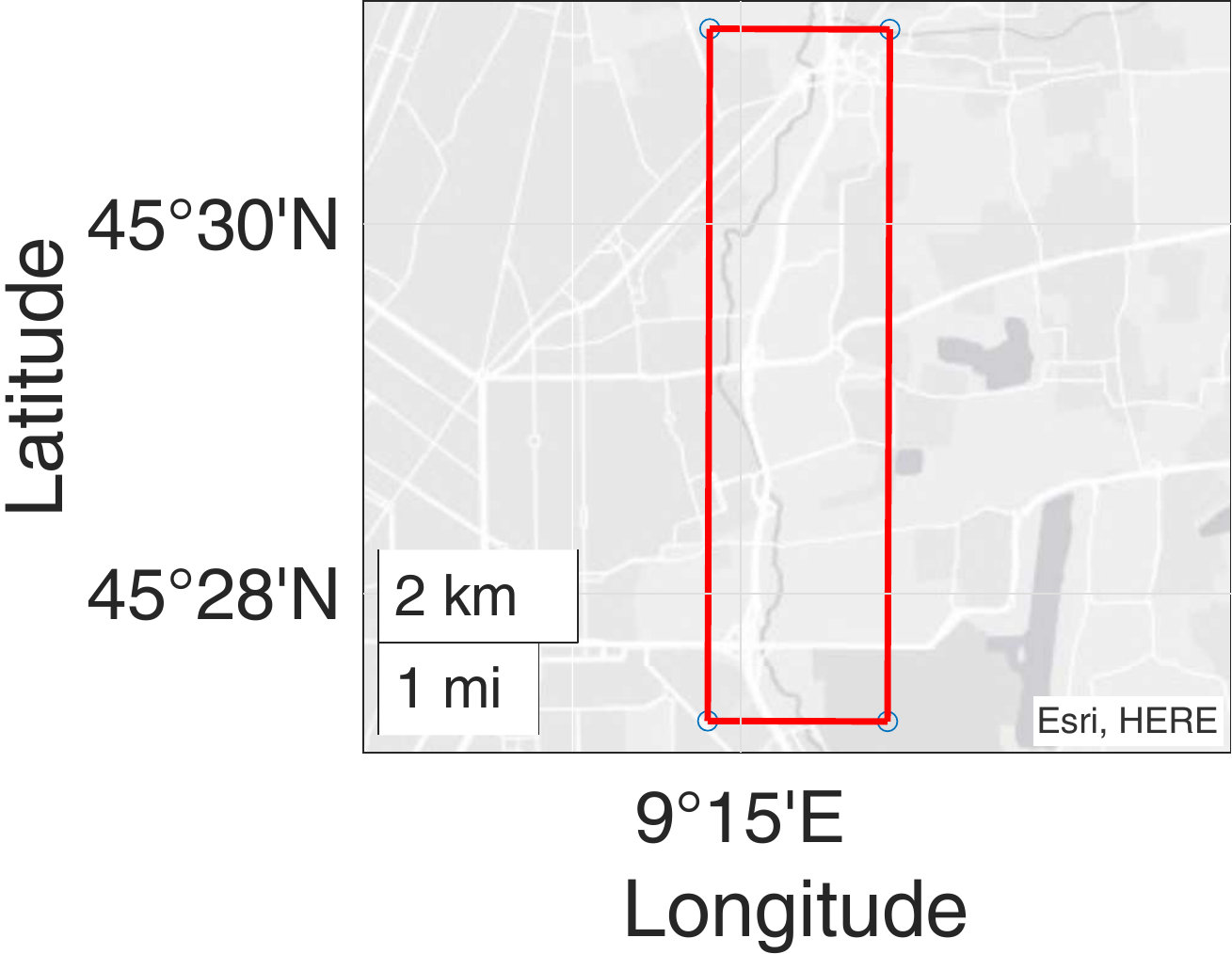}}
\caption{Simulated urban (a, b) and suburban (c) scenarios.}
\label{fig:scenarios}
\end{figure*}

We consider three different scenarios of vehicular mobility as show in Fig.~\ref{fig:scenarios}, namely, a road intersection, a roundabout in the area of the city of Milan, Italy, and a 14 km long stretch of sub-urban highway in the surroundings of the city. 
The geographical maps and objects (e.g., foliage, walls, and buildings) coordinates are taken from OpenStreetMap~\cite{osm}. 

The two urban areas have been chosen with the aim of covering different type of Line-of-Sight (LoS) blockage condition. So as to investigate the impact of different AoA/AoD statistics. 
The first area (Fig.~\ref{fig:intersection}) is characterized by a series of road intersections with an high building density, while the second one (Fig.~\ref{fig:roundabout}) is manly an open-space scenario with the presence of a wide roundabout, so that V2V LoS communications are made possible from different directions. The third area (Fig. \ref{fig: highway}), instead, is representative of high-speed mobility with mainly straight vehicle motion.

\subsection{Mobility and Localization}
The SUMO software~\cite{SUMO2018} simulates vehicular mobility. The driving behavior depends on the vehicle's type, length, height, and width. Therefore, different cars are considered, like passenger cars, motorcycles, taxis, and emergency ones. 

The road network topology is given as input from OpenStreetMap, together with the mobility parameters. For each vehicle, SUMO provides the corresponding vehicle position (in the form of latitude and longitude coordinates), speed, and direction of motion (i.e., heading). 
It is possible to configure the vehicular density, maximum velocity, driving behavior, and other mobility parameters.

We assume each vehicle to be able to measure its instantaneous position $\mathbf{p}_v(t)$ over time $t$ in a bidimensional (2D) space, e.g., through a GPS receiver, modelled as
\begin{equation}
    \mathbf{p}_v(t) = \mathbf{p}_{\text{SUMO}}(t) + \mathbf{e}_{v}(t) \,,
\end{equation}
where $\mathbf{p}_{\text{SUMO}}(t)$ contains the UTM\footnote{Universal Transverse Mercator} coordinates (obtained from converting the SUMO outputs) and $\mathbf{e}\sim \mathcal{N}(\mathbf{0}, \sigma_p^2\mathbf{I}_2)$ is the measurement error, with its standard deviation $\sigma_p$~\cite{Brambilla2020sensors}.

\subsection{Millimeter Wave Channel Model}

\begin{figure}[!t]
    \centering
    \includegraphics[width=0.7 \columnwidth]{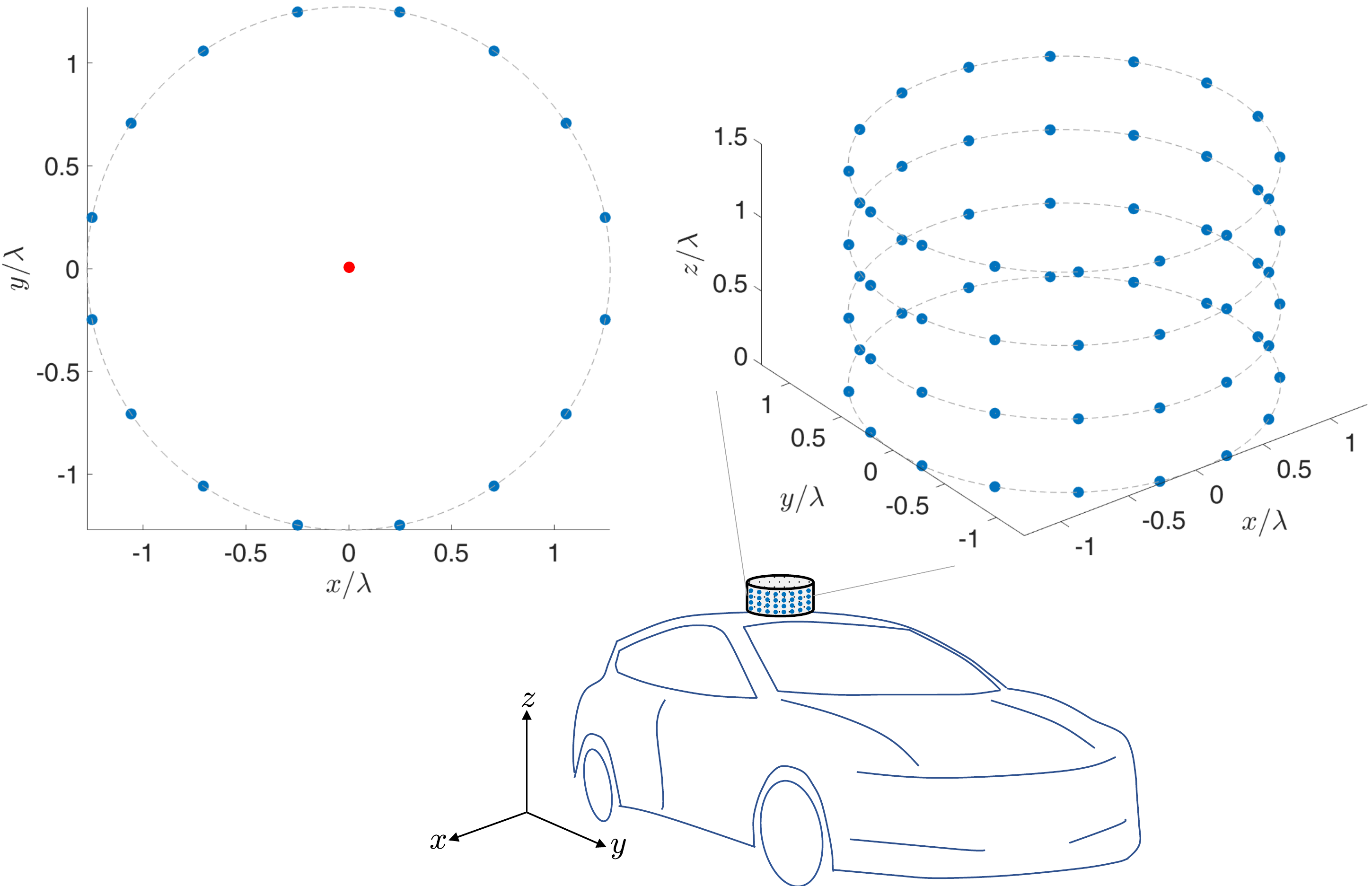}
    \caption{Antenna array model: left, UCA of $16$ antenna elements, right, cylindrical array of $4$ rings of UCAs. The reference of the array is the point (0,0,0), highlighted in red on the left.}
    \label{fig:array}
\end{figure}

The mmWave channel has been well studied at some typical frequency bands, such as the $26/28$,  $32$, $38/39$, $60$, and $73$ GHz bands. Compared to sub-$6$ GHz bands, mmWaves have very different channel propagation characteristics, such as the high path loss (PL) and high penetration loss. Thus, directional antennas with beamforming techniques are necessary to communicate at a reasonable range. However, this introduces a whole set of technological challenges, such as the frequent beam blockage and misalignment, which are exacerbated in a highly dynamic scenario as V2V. 
To better capture all these aspects, the mmWave channel is simulated with GEMV$^2$~\cite{GEMV2}. This makes use of real-world data (locations, dimensions, 3D topology) to determine propagation conditions between vehicles. It evaluates the large-scale channel components with a deterministic approach and the small-scale ones with a geometry-based stochastic approach accounting for surrounding objects. The PL model is computed based on the 3GPP guidelines in~\cite{14rel}. 
The LoS condition is assessed geometrically exploiting the possible intersection between the outlines of buildings, foliage, or vehicles and the beams of the communications link pair.  
First order reflections, which in mmWave range can carry a non-negligible amount of power, are considered. 
Reflections are computed geometrically by GEMV$^2$ \cite{GEMV2}, which models the multipath propagation.
In particular, we account only link where LoS condition (no blockage) is present~\cite{GEMV2}. This leads to the PL computation for the direct component as follows~\cite{14rel}
\begin{align}
   \text{PL}_{\text{LoS}} = 32.4 + 20\log_{10}d_{PL} + 20\log_{10}f \,,
\end{align}
where $d_{PL}$ is the distance between V-Tx and V-UE, and $f$ the carrier frequency.
For each pair of vehicles (V-Tx and V-UE), and for each path/ray direct and/or reflected $p=1, 2,\cdots, P$, GEMV$^2$ computes  the azimuth and elevation Angle of Departures (AoD) $(\vartheta^{T}_{p}, \phi^{T}_{p})$, the azimuth and elevation Angle of Arrival (AoA) $(\vartheta^{R}_{p}, \phi^{R}_{p})$ (see Fig.~\ref{fig:reference}), and the complex amplitude $\alpha_{p}$ which accounts for the PL, time of delay, and Doppler. It follows that the channel matrix can be written as
\begin{align} 
    \mathbf{H} = 
    \mathbf{A}_{R}
    \left(\boldsymbol{\vartheta}^R,\boldsymbol{\phi}^R \right) 
     \mathbf{D} \,
    \mathbf{A}_{T}^{\mathrm{H}} 
    \left(\boldsymbol{\vartheta}^T, \boldsymbol{\phi}^T \right)  \,,
\label{eq:channel_matrix_compact}
\end{align}
where $\mathbf{D} \in \mathbb{C}^{P \times P} = \mathrm{diag}\left(\alpha_1, \dots, \alpha_P \right)$ is diagonal matrix that collects all the channel $P$ complex amplitudes, 
$\mathbf{A}_T \left(\boldsymbol{\vartheta}^T, \boldsymbol{\phi}^T \right) = \left[\mathbf{a}_T(\vartheta^{T}_{1}, \phi^{T}_{1}),\dots,\mathbf{a}_T(\vartheta^T_P, \phi^T_P)\right] \in \mathbb{C}^{N_a\times P}$
and $\mathbf{A}_R\left(\boldsymbol{\vartheta}^R, \boldsymbol{\phi}^R\right)$
are the two matrices identifying the Tx and Rx \textit{beam spaces}, which are composed of the set steering vectors
$\mathbf{a}_{T}(\vartheta_{p}^T,\phi_{p}^T)$ and $\mathbf{a}_{R}(\vartheta_{p}^R,\phi_{p}^R)$ that includes the  response of each antenna element $a_{mn}(\vartheta,\phi)$. We assume that each vehicle is equipped  with a cylindrical antenna array on its rooftop as shown in  Fig.~\ref{fig:array}, with $N_a\!=\!N_v N_c\!=\!64$ antennas, where $N_v=4$ is the number of uniform circular arrays (UCA) each with $N_c=16$ antenna elements. 
The response of element $m$ in the $n$th UCA of the V-Tx/UE array is
\begin{align}\label{eq:ula_resp}
    a_{mn}(\vartheta,\phi) =  e^{\,j\frac{2\pi}{\lambda} r\cos\phi\cos(\vartheta-\vartheta_{m})}
    \cdot 
    e^{\,j\frac{2\pi}{\lambda} d\cdot(n-1)\sin\phi}  \,,
\end{align}
where $\lambda$ is the wavelength, $d$ is the element spacing that is set as $\lambda/2$ (also among two different UCA elements), $\vartheta$ is the azimuth, $\phi$ is the elevation angle, $r$ is the UCA radius and
\begin{align}
    \vartheta_{m} = (2m - 1)\cdot\frac{\pi}{N_{c}} 
\end{align}
is the angular position of $m$-th element. 
\input{system/referenceSystem}

\subsection{Evaluation of the SNR}

Assuming perfect synchronization, the discrete-time received signal vector at each time instant can be written in the following matrix form
\begin{align} \label{eq: received signal}
    \mathbf{y} = \mathbf{H\,x+n} = \mathbf{H\,f}\,s+\mathbf{n} \,,
\end{align}
where $s$ is the complex symbol taken from $M$-QAM constellation, with null mean and variance $\sigma^2_s$, at the output of the modulator and at input of the V-TX beamformer $\mathbf{f} \in \mathbb{C}^{N_a\times 1}$, vector $\mathbf{x} = \mathbf{f}\, s$ is the transmitted signal, $\mathbf{n} \sim \mathcal{CN}\left(\mathbf{0}, \mathbf{\sigma^2}_n\right)$ is the additive noise, while $\mathbf{H} \in \mathbb{C}^{N_a \times N_a}$ is the MIMO channel matrix given in (\ref{eq:channel_matrix_compact}). 

At the receiver side, if $\mathbf{w} \in \mathbb{C}^{N_a \times 1}$ is the V-UE beamformer, the the received symbols $\Tilde{s}$ can be computed as
\begin{align}\label{eq:detect}
    {\Tilde{s}} = \mathbf{w}^{\text{H}}\textbf{y} = \mathbf{w}^{\text{H}} \,\mathbf{H} \, \mathbf{f} \, s + \mathbf{w}^{\text{H}} \mathbf{n}\,,
\end{align}
where $(\cdot )^{\text{H}}$ denotes the Hermitian operator.

In the case of perfect beam alignment, the beamforming vectors at V-TX and V-UE sides are defined as, respectively,
\begin{align}\label{eq:beamformer}
    \mathbf{f} &= \frac{1}{\sqrt{N_{v}N_{c}}}\cdot\mathbf{a}_T\left(\vartheta^T_m, \phi^T_m\right)  \, , \nonumber\\
    \mathbf{w} &= \frac{1}{\sqrt{N_{v}N_{c}}}\cdot \mathbf{a}_R\left(\vartheta^R_m,\phi^R_m\right)\,,
\end{align}
where $\mathbf{a}_T\left(\vartheta^T_m, \phi^T_m\right)$ and $\mathbf{a}_R\left(\vartheta^R_m,\phi^R_m\right)$ are computed based on \eqref{eq:ula_resp}  and $(\vartheta^T_m, \phi^T_m)$ and $(\vartheta^R_m, \phi^R_m)$ are the AoD/AoA of the direct LoS path.
\begin{figure*}[b]
\centering 
\subfloat[Instantaneous evolution \label{fig:snr_track}]{ \includegraphics[width=0.45\columnwidth]{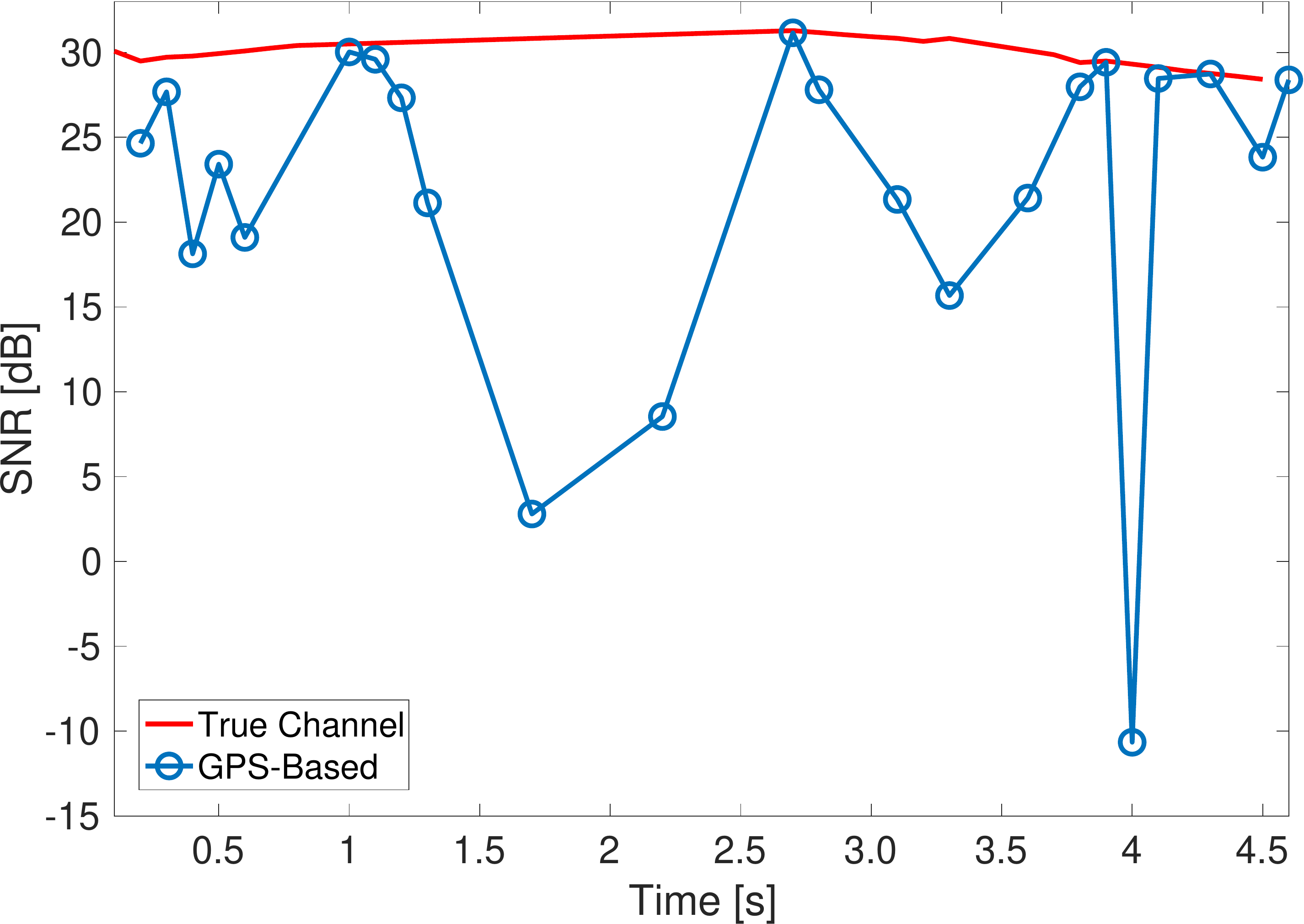}}
\subfloat[ECDF \label{fig:snr_gps_ideal}]{\includegraphics[width=0.45\columnwidth]{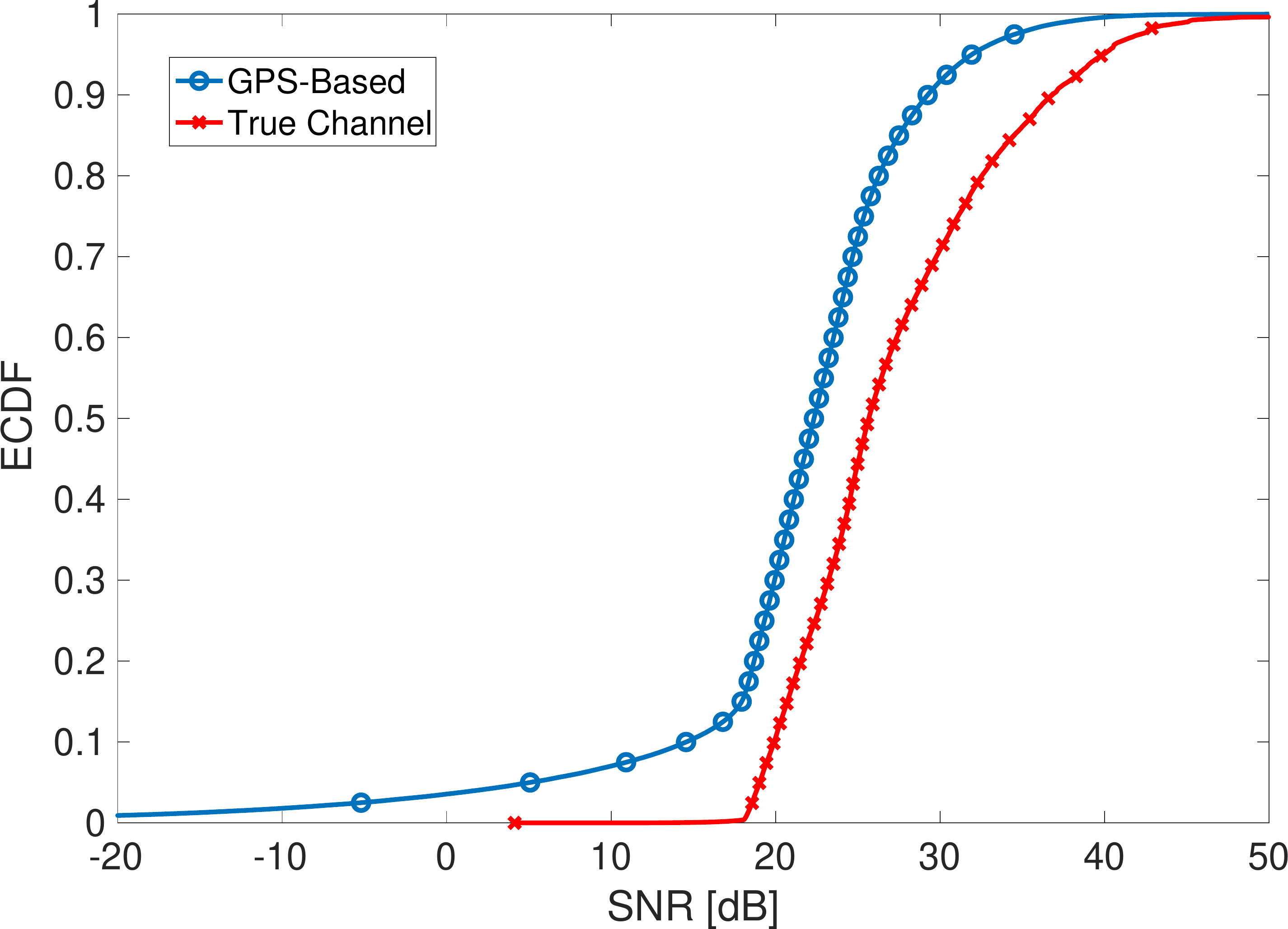}}
\caption{SNR comparison for LoS V2V sidelink communications: True Channel versus GPS-Based.(a) example of time-series evolution, (b) ECDF of the aggregated data.}
    \label{fig:snr_gps_ideal_all}
\end{figure*}

Finally, the overall system SNR is computed as
\begin{align}\label{eq:SNRdef}
    \text{SNR} &= \frac{\abs{\mathbf{w}^\mathrm{H} \,\mathbf{H} \, \mathbf{f}}^2 \,\sigma_s^2}{N_a \sigma^2_n}  \\
    &= \frac{\sigma_s^2}{\sigma_n^2} \, \frac{\abs{\mathbf{w}^\mathrm{H}  \mathbf{A}_R\left(\boldsymbol{\vartheta}_a, \boldsymbol{\phi}_a\right) \mathbf{D}\, \mathbf{A}_T^{\text{H}}\left(\boldsymbol{\vartheta}_d, \boldsymbol{\phi}_d\right) \mathbf{f}}^2}{N_a} \nonumber \,,
\end{align}
where $\abs{\cdot }$ is the absolute value operator.

Figure~\ref{fig:snr_gps_ideal_all} depicts an example of SNR evolution (Fig.~\ref{fig:snr_track}) and the Empirical Cumulative Distribution Function (ECDF) (Fig.~\ref{fig:snr_gps_ideal}) that are obtained with the beamforming computed based on the V-TX and V-UE positions, i.e., on the direct link. The continuous line refer to perfect knowledge of the positions with no error, i.e., $\mathbf{p}_v(t) = \mathbf{p}_{\text{SUMO}}(t)$, while the blue line assume the position known with an error of $\sigma_p = 4$ m~\cite{Brambilla2020sensors}. From these results, we can state that even the optimal LoS condition can become not favourable if localization errors occur. 

\subsection{Codebook-Based Beam-Sweeping}

As discussed in Section~\ref{sec:IAin5GNR}, during the IA, the V-TX performs a beam-sweeping procedure where it transmits the S-SSB in a set of predefined directions $\boldsymbol{\vartheta} \in \left\{\vartheta_i\right\}_{i=1}^{K_a}$, with $\vartheta_i \in [0, 2\pi)$ and $K_a$ is the maximum number of sweeps in azimuth and $\boldsymbol{\Phi} \in \left\{\phi_j  \right\}_{j=1}^{K_e}$ with $\phi_j \in [-\pi/2, \pi/2]$ and $K_e$ is the maximum in elevation. This set constitutes the beamforming codebook, and it must cover all possible angles. The codebook \textit{depth} is $K_c = K_a K_e$, and it represents the total number of beam-sweeps that the V-TX has to perform. A high depth value increases the performances (e.g., higher SNR, lower misdetection rate) and the system complexity (i.e., delay and overhead of the IA procedure). Thus, the codebook design must consider a trade-off when defining $K_c$. 

Since we target V2V systems, all vehicles have a similar height, therefore we neglect the elevation angles (i.e., $K_e = 1$) and we focus only on azimuth ones \cite{LinsICC-c21}. The codebook depth is 
\begin{equation}
    K_c = K_a = \ceil* {\frac{2\pi}{\vartheta_q}} \,,
\end{equation}
where $\vartheta_q$ is the quantization threshold. To improve the system performance and overcome limitations of uniform codebooks, Section~\ref{sec:BSS} will address azimuth angles' selection and quantization strategy for beamforming codebook design.

\section{Beam Selection Schemes}\label{sec:BSS}

\begin{figure*}[!t] 
\centering 

\subfloat[Intersection \label{fig:sub1}]{\includegraphics[width=0.19\columnwidth]{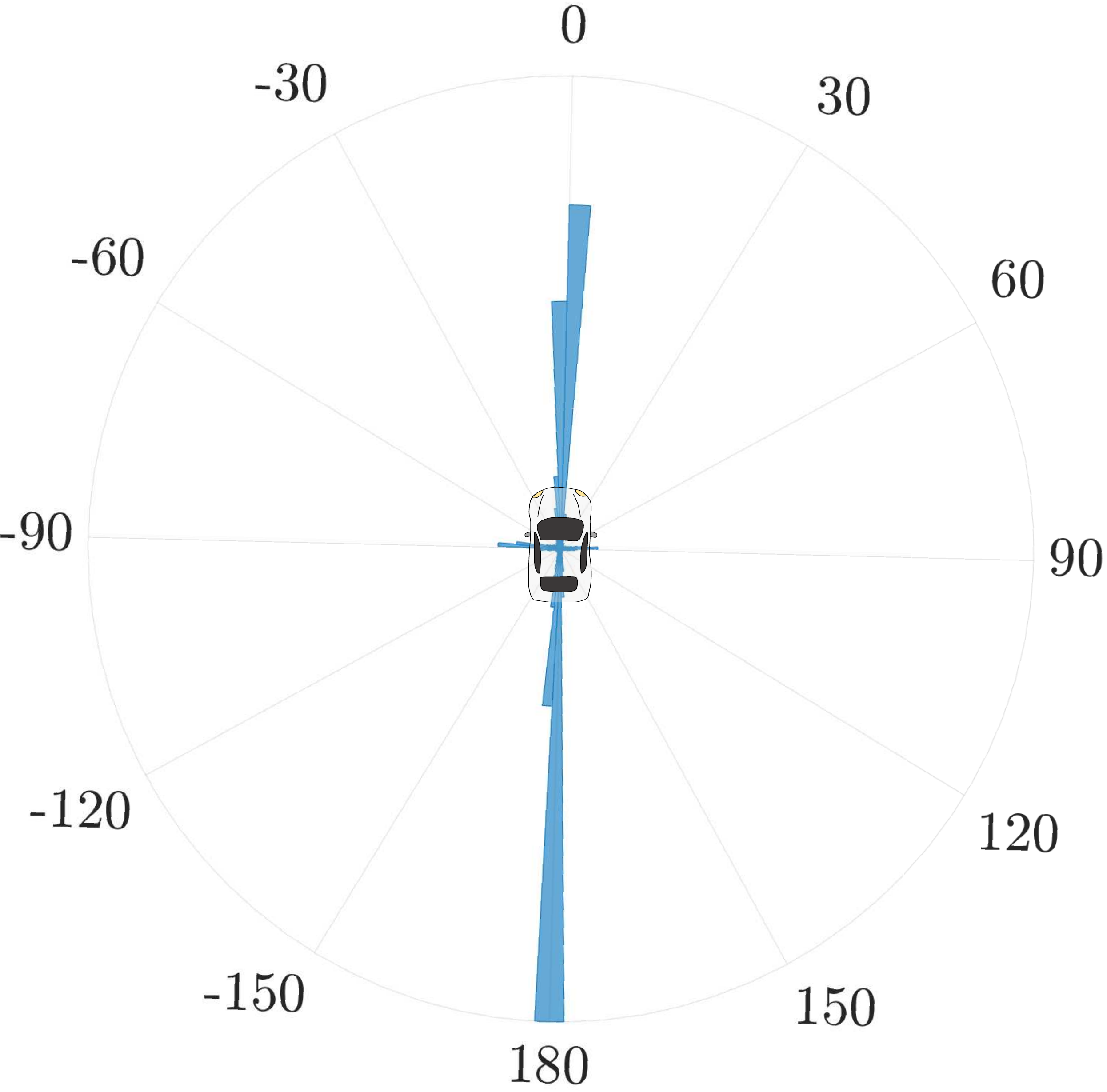}}
\subfloat[Roundabout \label{fig:sub2}]{ \includegraphics[width=0.19\columnwidth]{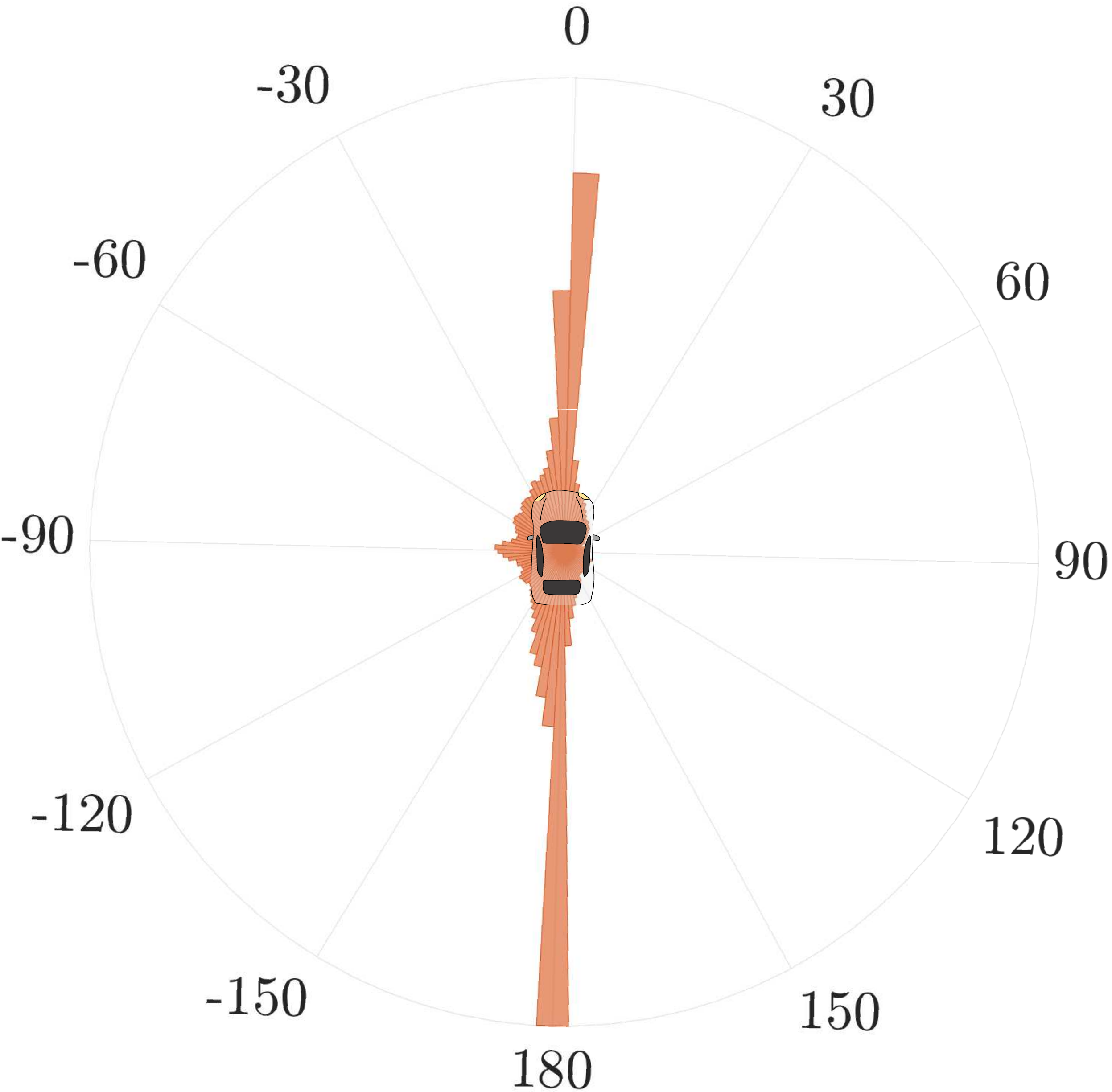}}
\subfloat[Highway\label{fig:sub3}]{\includegraphics[width=0.19\columnwidth]{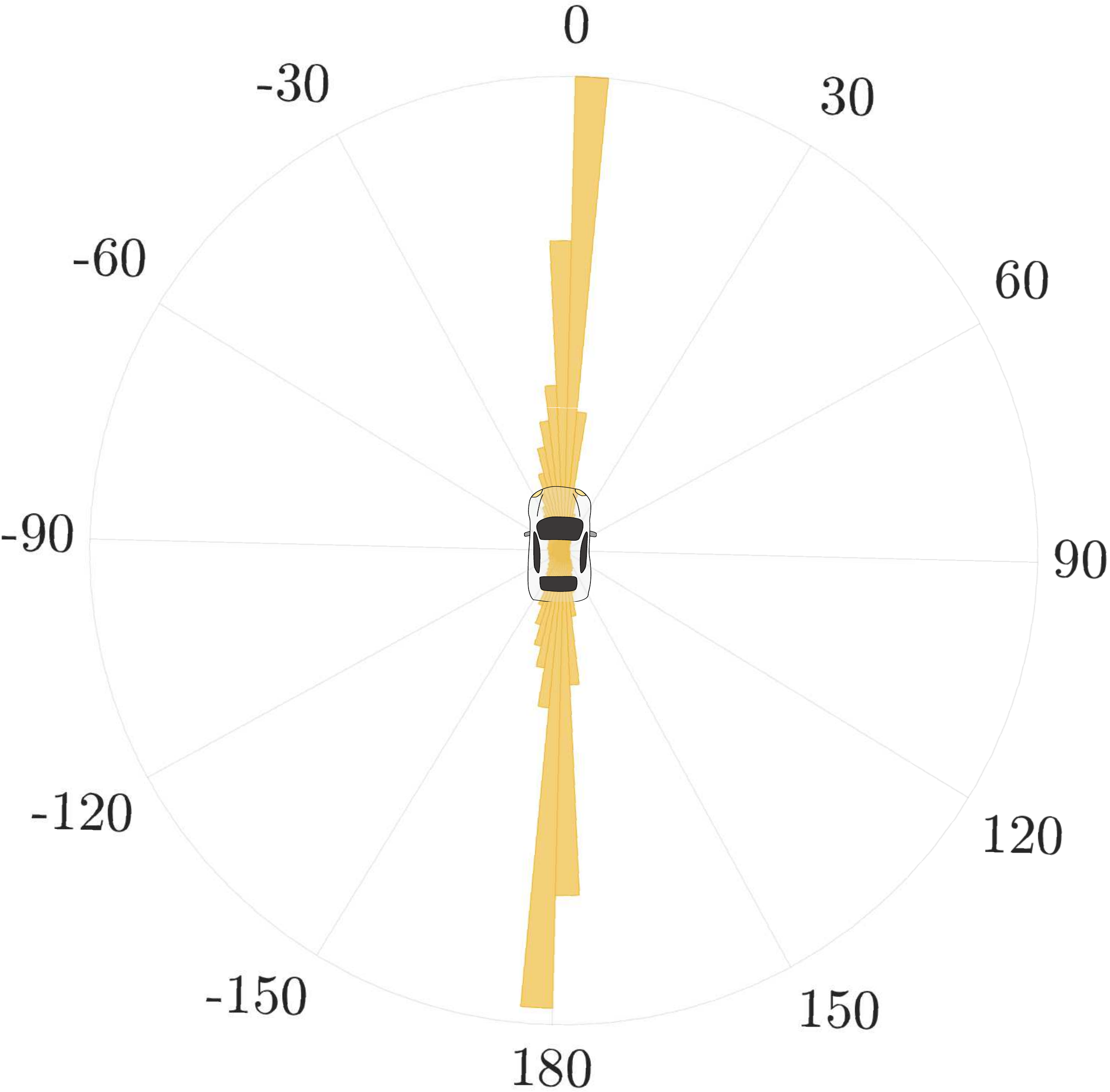}}
\hspace{2mm}
\subfloat[Azimuth probability densities \label{fig:sub4}]{\includegraphics[width=0.4\columnwidth]{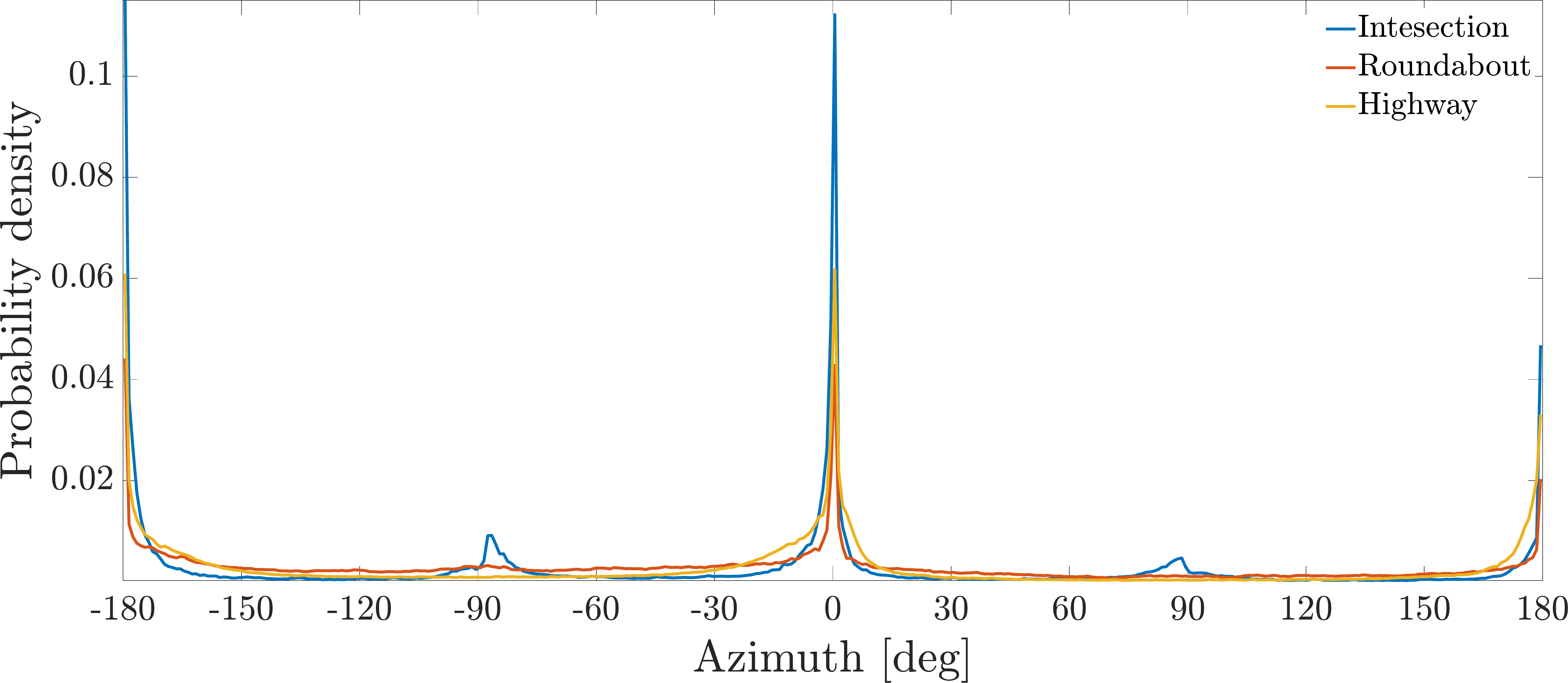}}
\caption{Polar histogram of azimuth angles for the separate cases of (a) intersection, (b) roundabout and (c) highway scenarios.  (d) Probability densities of azimuth angles for all the considered scenarios.}
\label{AoA}
\end{figure*}

In this section, we address the problem V2V beam selection and alignment in azimuth\footnote{Note that in our previous research~\cite{LinsICC-c21} we demonstrated how variation in the elevation angle can be considered as negligible.}. This is a key aspect for a reliable V2V communications, in fact, in case of incorrect beam alignment the SNR in~\eqref{eq:SNRdef} rapidly decreases. 

Beam directions are defined in a spatial reference system with origin $(x_{0},y_{0},z_{0})$, where $(x_{0},y_{0})$ coincides with the vehicle's antenna array position in the $xy$-plane, while $z_{0}\! =\! h_{v}+0.1$ m, where $h_{v}$ is the vehicle height. It is worth noting that for these schemes, the beam direction (i.e., AoA/AoD) is measured with respect to the vehicle heading, as indicated in Fig.~\ref{AoA}, with positive angles as clockwise and negative angles counter clockwise.

\subsection{{\textit{Position-Assisted Schemes}}}
Position-assisted approaches rely on the information of position, which turns out to be of high relevance and allows to reach the correct alignment in few attempts. However, this information might not be available since it needs to be retrieved through a different interface, e.g., from sensing or signaling in FR1~\cite{LiYuFukTraSak:J21,Brambilla2020sensors,GarRogMarMonKouSpaZho:J19}. Moreover, the accuracy should be taken into account as it may decrease the received SNR a showed in Fig.~\ref{fig:snr_gps_ideal_all} for the case of $\sigma_p = 4$ m ~\footnote{Results have been obtained in the urban roundabout scenario of Fig.~\ref{fig:roundabout}, with the same simulation parameters as in Sec.~\ref{sec:performanceevaluation}}. 
The goal of the SNR analysis is to show that an inaccurate vehicle position information can lead to severe degradation of SNR, as indicated by the example of the time-series in Fig.~\ref{fig:snr_track}, as well as from an aggregated analysis that takes into account the ECDF of the SNR in Fig.~\ref{fig:snr_gps_ideal}.

The position-assisted schemes can be categorized as follows:
\begin{itemize} [wide]
   \item {\textit{GPS-assisted beam selection}} (e.g.,~\cite{Hanzo:J2019,Va:J2018,Nordio:J2020}): it relies on the reciprocal knowledge of current position information between  V-TX and V-UE, e.g., individually obtained from GPS systems or from network infrastructure (if they are in coverage), and shared through high-level single hop Cooperative Awareness Messages (CAM) ~\cite{CAMstandard}. This cooperative approach allows the IA beam searching phase to start from a candidate \textit{position-optimal} beam, which is determined starting from the mutually-received signaled position information. In case of failure (e.g., due to poor position estimate or blockage), a left-right jumping search is iterated by the V-UE until a match is found. 
   \item {\textit{GPS-assisted adaptive beam selection}} (e.g.,~\cite{Va:C2016,BrambillaICC2020,Feng:C2019,lms_smart_array}): it can be seen as a variant of the previous cooperative approach that tries to fasten the GPS-assisted beam selection. While the latter performs a jumping search around the position-aided starting beam, this adaptive version avoids unnecessary trials by adopting Least Mean Square (LMS) technique, which is an iterative optimization algorithm based on the gradient method~\cite{lms_smart_array}, where the test of successive beams tries to maximize the received power/beamforming gain. 
   
   The updating system is set as follows
    \begin{align}
        \vartheta_a^{k+1} = \vartheta_a^{k} + \eta^{k}\,\epsilon^k{b}^k \,,
    \end{align}
    with     ${b}^k$ = $ \abs{\mathbf{w}_k^{\text{H}}\,\mathbf{f}}  G_{max}^{-1}$,
    $\epsilon^{k}$ = $1-{b}^k$, and $\eta^{k}$ = $\text{sign}(\epsilon^{k-1}-\epsilon^{k})\,\mu^{k-1}$,
    where $k$ stands for the $k$th iteration, $G_{max}$ is the expected maximum gain according to the array design, ${b}^k$ is the observation with the corresponding normalized error $\epsilon^k$, and $\eta$ is the step-size. The value of $\vartheta^0$ is computed from the received position information (e.g., GPS data), while the step-size is set according to the trade-off between the required time of convergence and accuracy (e.g., $\eta^0=0.05$). 
    The conditions to break the updating are set according to the accepted accuracy error $\epsilon^k$ and the maximum number of beam of attempts (i.e., maximum $64$ trials).  
    
\end{itemize}   
   
\subsection{{\textit{Probabilistic Codebooks (PCBs)}}} \label{sec:PCB} 

The beamforming codebook is defined by a set of angles (only azimuth is considered) $\boldsymbol{\vartheta}$ that are sorted according to some policy. In 5G NR standard (Sec. \ref{sec:IAin5GNR}), without any prior knowledge, the set of azimuth angles is derived as
\begin{equation}\label{eq:uniformCB}
    \boldsymbol{\vartheta}_{\text{5G NR}} = \left\{0, \frac{2\pi}{N_{s}}, 2\frac{2\pi}{N_{s}}, \dots, (N_{s}-1)\frac{2\pi}{N_{s}} \right\} \,,
\end{equation}
where $N_{s}$ is the number of S-SSBs transmitted for each $160$ ms period. The V-Tx starts the beam-sweeping procedure with the beamformer $\mathbf{f}^0 = \mathbf{a}_T(0,0)$ and complete the procedure with the beamformer $\mathbf{f}^{K_c} = \mathbf{a}_T((N_{s}-1)2\pi/N_{s},0)$. This method assumes a uniform angular probability density function $p_{\vartheta}(\vartheta)$. However, since the LoS in V2V sidelink communications is conditioned to the surrounding constrains (e.g., road topology, buildings and foliage position/density), the AoA/AoD distribution is not uniform. Indeed, some directions are expected to be more likely than others as it can be observed in Fig.~\ref{AoA}. Thus, it is possible to rely on this knowledge to fasten the IA beam search over some directions, i.e., starting from those with the highest probability (see Fig. \ref{AoA}), and reducing the overall number of trials for the beam selection. 
Note that a  probabilistic codebook takes into account propagation condition and blockages  typical of a given environment, thus being substantially different from the GPS-based geometrical approach, as explained in \cite{ZenXu:J21}.

The main goal of this section is to show how the probabilistic codebook can be obtained. We have investigated two different methods, one requiring a training phase to acquire information on the most prevalent communication directions, and the other one leveraging on the Hough Transform (HT) tools, an image processing tools that allows to effectively extract building outlines from a digital-map of the environment~\cite{rs11141727}.  The details of these two methods are in the following.

\begin{algorithm}[!t] 
\begin{small}
\caption{Simulation-based PCB}
\label{alg:PCB-simulated}
\begin{algorithmic}[1]
\Require{vehicle position $\mathbf{p}_v(t_k)$ at $k$th time instant}
\Ensure{$\boldsymbol{\vartheta}$ for beam-sweeping in IA }
\State V-Tx sends its $\mathbf{p}_v(t_k)$ to the eNB/gNB
\State eNB/gNB determines from  $\mathbf{p}_v(t_k)$ the current vehicle quadrant $q$th 
\State eNB/gNB extracts the latest $\boldsymbol{\vartheta}_{q}$, i.e., $\vartheta$ sorted based on the probability density $p_{\vartheta}(\vartheta)$ as in Fig.~\ref{AoA}, and reports it to V-Tx 
\State V-Tx starts searching from the most probable angle until the match with V-UE is found
\State V-Tx send backs the matching AoA/AoD 
\State eNB/gNB updates the database and the $\boldsymbol{\vartheta}_{q}$
\end{algorithmic}
\end{small}
\end{algorithm}

\begin{itemize} [wide]
  \item {\textit{PCB-based on a training phase}}: the PCB-based beam selection has been implemented according to the following two steps procedure: \textit{i)} the area is divided into sub-regions (or quadrants) with fixed size/footprint and shape. For each quadrant, a statistical analysis of AoA/AoD distribution of the beam pointing is learned over multiple vehicle passages in a given $q$ area. The learned PCBs can be stored in the cloud with their related geo-location, or in a channel knowldege map \cite{ZenXu:J21}; \textit{ii}) the vehicles can download the specific PCB based on their position (for autonomous vehicles, multiple PCBs can be downloaded based on the planned trajectory) and use it for the a fast IA with the vehicles in the nearby. The implemented pseudocode is reported in Algorithm~\ref{alg:PCB-simulated}. Differently of the position-assisted schemes, here vehicles only need to know their own positions and no type of information to be exchanged with other vehicles is needed. The choice of the quadrant size should represent a trade-off with respect to overhead (small quadrants mean more codebook updates) and position accuracy (localization error induce a wrong choice of the codebook, i.e., the codebook of a different quadrant is chosen).
  A possible solution is that each quadrant should be chosen such that it coincides with a specific road segment (i.e., crossroad, T-junction, straight road) or environment type (i.e., regular grid-like, highway) that presents a peculiar AoA/AoD distribution.
  The main drawbacks of this approach are strictly related to the need of a training phase. In fact, privacy impairments can arise in the data collection/sharing phase. Moreover, the overhead due to codebook updates must be considered.
 Motivated by these reasons, we decide to investigate another possible solution to determine the codebook without relying on a training phase and sensitive shared information. 
  \item {\textit{PCB-based on digital Map}}:
  the key idea behind the HT is to determine the most recurrent straight-lines with their own rotation based on a voting scheme in a parameter space~\cite{rs11141727}. 
  The transformation consists in mapping all the points of an image $\mathbf{IM}$\footnote{The matrix dimension depends on the image format.} from the $xy$-plane to corresponding sinusoidal dual $\rho\vartheta$-plane as~\cite{rs11141727}
  \begin{align} \label{eq: HT}
      \rho= x \cos{\vartheta}+ y \sin{\vartheta} \,, \quad \text{ with } \vartheta \in [-90^\circ, 90^\circ) \,,
  \end{align}
  fixing $x$ and $y$, $\rho$ and $\vartheta$ represent respectively the distance from the origin and the orientation of all the straight lines passing through the point $(x,y)$. In a discretized representation of the $\rho\vartheta$-plane, the intersections of different sinusoids are represented by an accumulator matrix $\mathbf{H}_{HT}$. 
\end{itemize}

\begin{figure}[!t] 
\centering
\subfloat[Digital Map \label{fig:digital map}]{
\includegraphics[width=0.45\columnwidth]{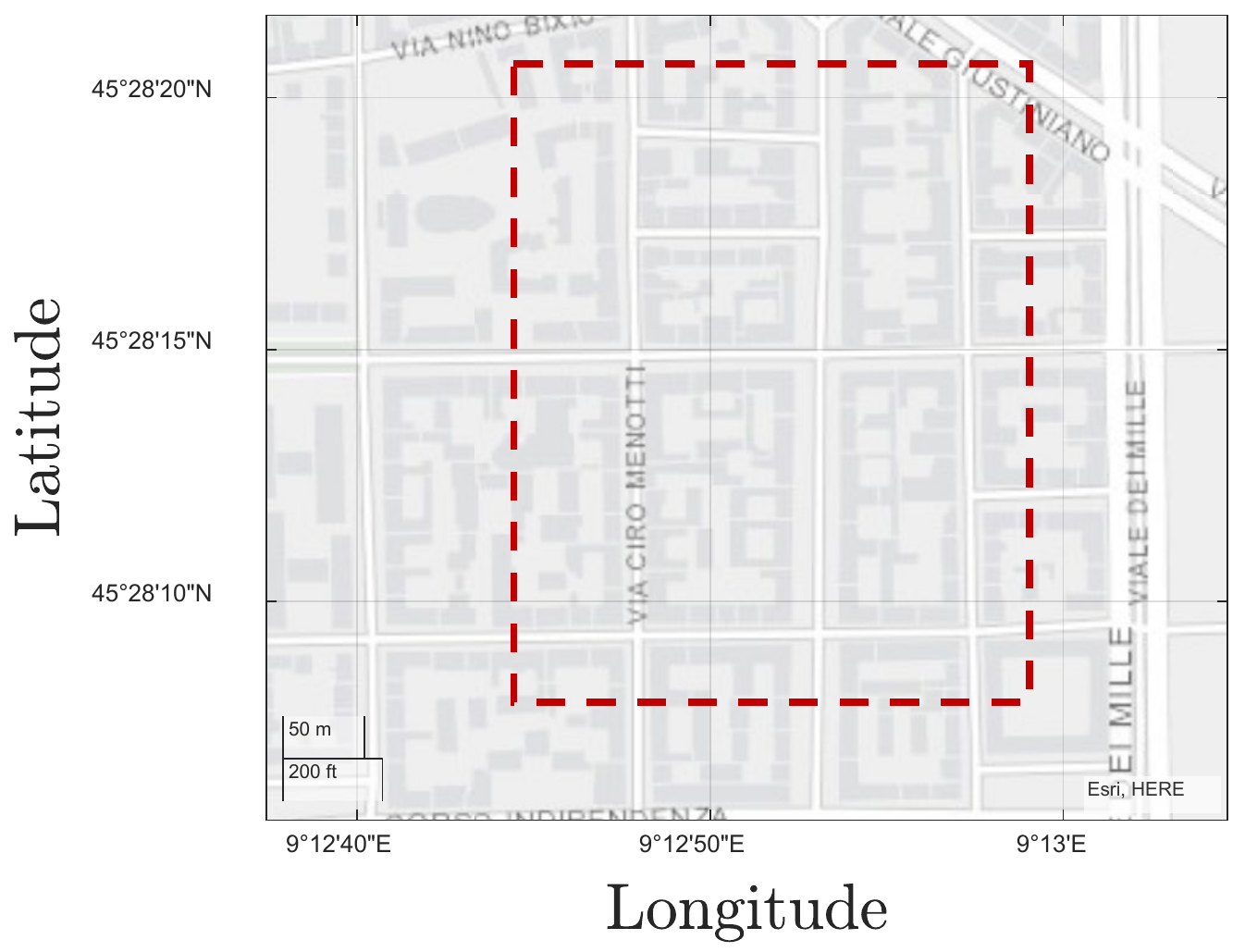}}\hspace{5mm}
\subfloat[Edges Extraction \label{fig:digital map BW}]{ \includegraphics[width=0.45\columnwidth ]{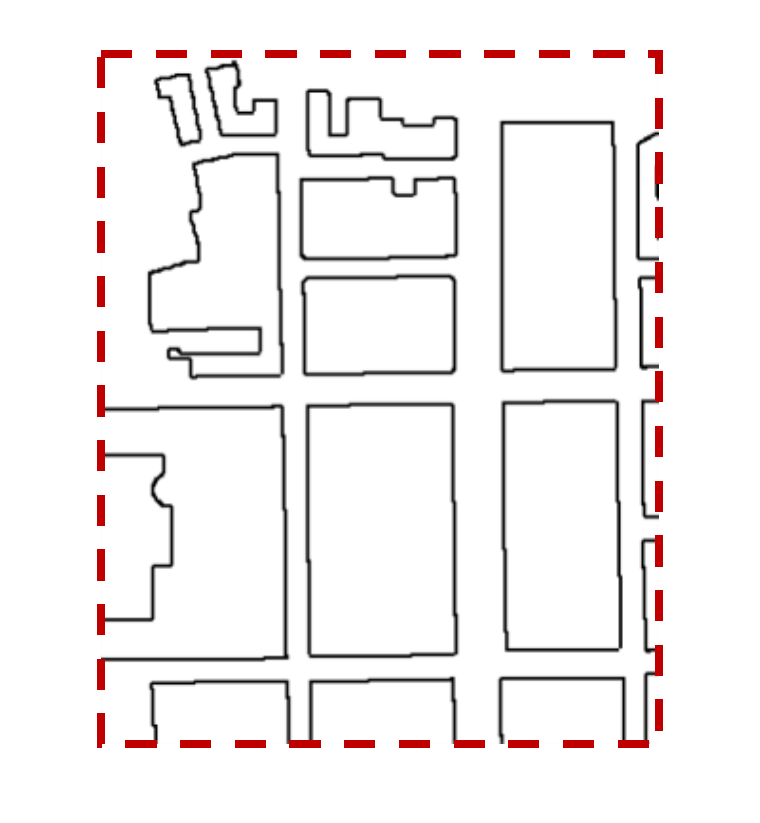}
}
\\
\subfloat[Hough Transform Output \label{fig:HT results}]{ 
\includegraphics[width=0.45\columnwidth ]{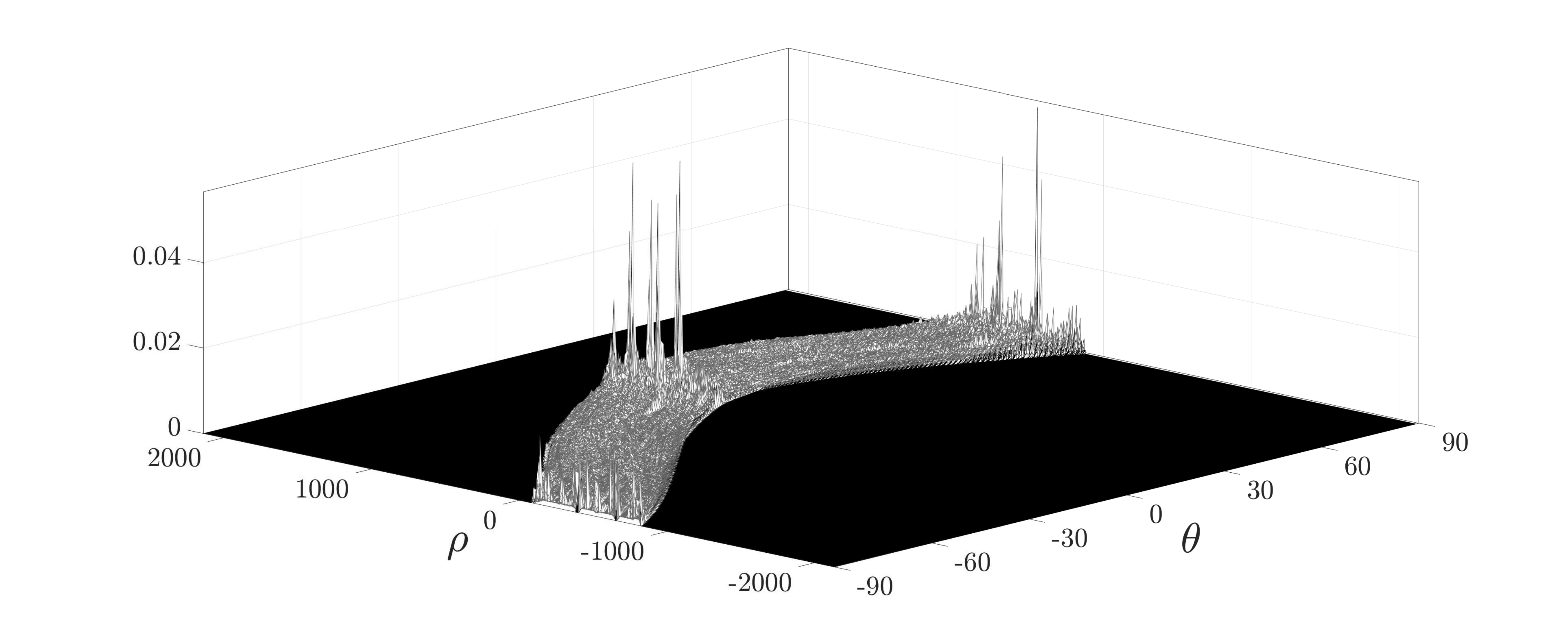}}\hspace{5mm}
\subfloat[Azimuth probability densities \label{fig:pdfs_comparison}]{\includegraphics[width=.45\columnwidth ]{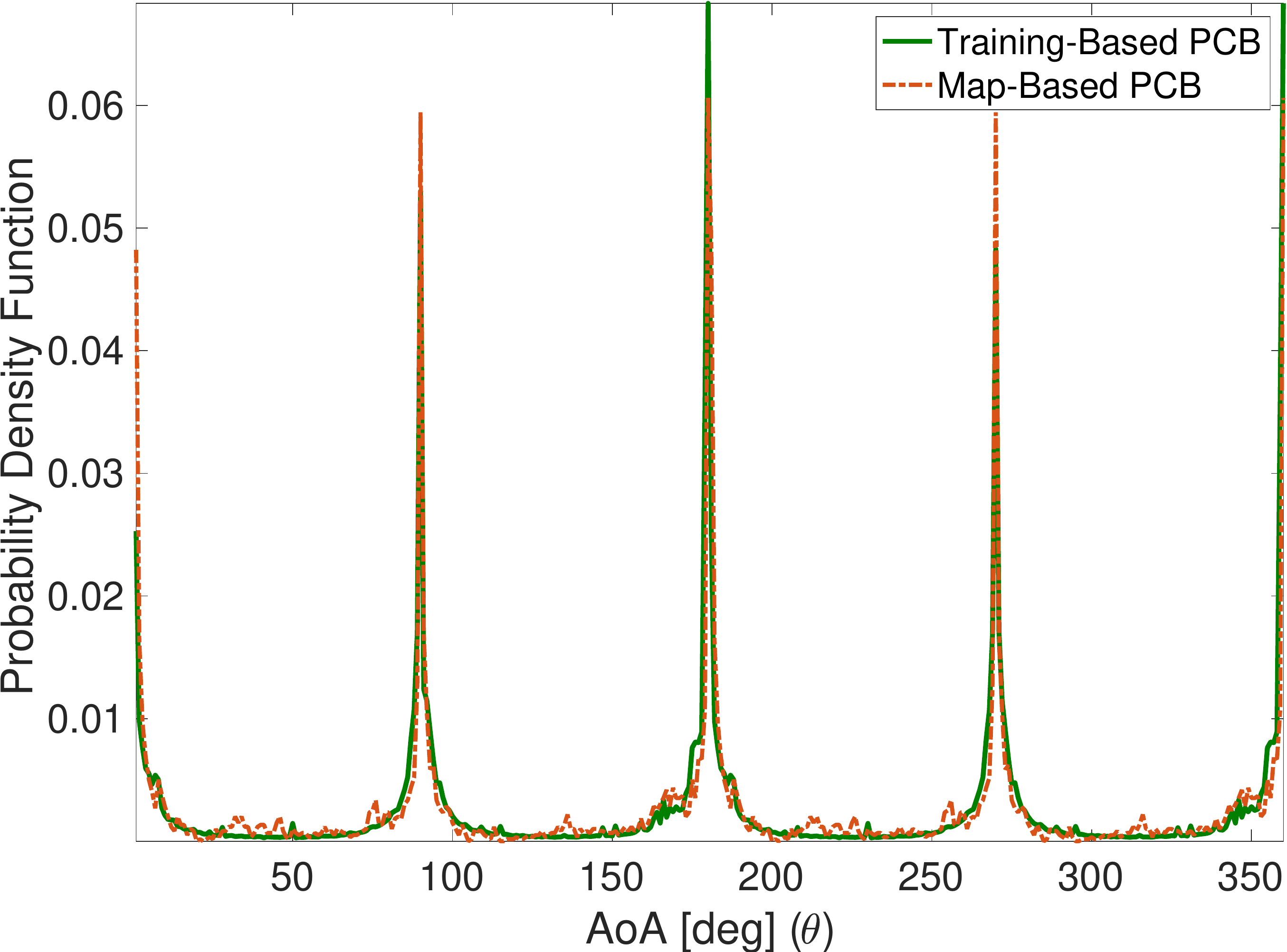}}
\caption{Hough Transform-based PCB steps for intersection.}
\label{fig:HT steps}
\end{figure}

   The implemented pseudocode to extract the AoAs/AoDs probability density using the HT is reported in Algorithm~\ref{alg:PCB-HT}, where we list the required  steps to get the outputs in Fig. \ref{fig:HT steps}.
\begin{algorithm}[!t] 
\begin{small}
\caption{Map-based PCB extraction }
\label{alg:PCB-HT}
\begin{algorithmic}[1]
\Require{e-Map $\mathbf{IM}_{e-maps}$ of intended path}
\Ensure{$\boldsymbol{\vartheta}$ for beam-sweeping in IA }
\State obtain an image $\mathbf{IM}_{e-maps}$ of road/buildings contours as Fig.~\ref{fig:digital map}~\cite{osm} 
\State apply high pass Prewitt Filter $\mathbf{IM}_{bin}$=PF($\mathbf{IM}_{e-maps}$) 
\State $\mathbf{IM}_{bin}$ is the binary image of edges as in Fig.~\ref{fig:digital map BW} 
\State apply HT in (\ref{eq: HT}), $\mathbf{H}_{HT}$=HT($\mathbf{IM}_{bin}$) as in Fig.~\ref{fig:HT results}
\State  compute  $p_{\vartheta}(\hat{\vartheta})$\,=\,$ (\underset{\rho}{max}( \mathbf{H}_{HT}) - \underset{\vartheta}{min} ( \underset{\rho}{max} (\mathbf{H}_{HT})))/\text{sum}(\mathbf{H}_{HT})$ 
\State $\hat{\vartheta}$ is mapped into AoA/AoDs
\Statex \,\,\,$\boldsymbol{\hat{\vartheta}}$\,=\,$\hat{\vartheta}$\,+\,$[0^\circ \, 180^\circ]$
\Statex \,\,\,rotate $\mathbb{\hat{\vartheta}}$ according to angular the reference system 
 \end{algorithmic}
\end{small}
\end{algorithm}

\begin{figure}[!b]
    \centering
    \includegraphics[width=.5\columnwidth]{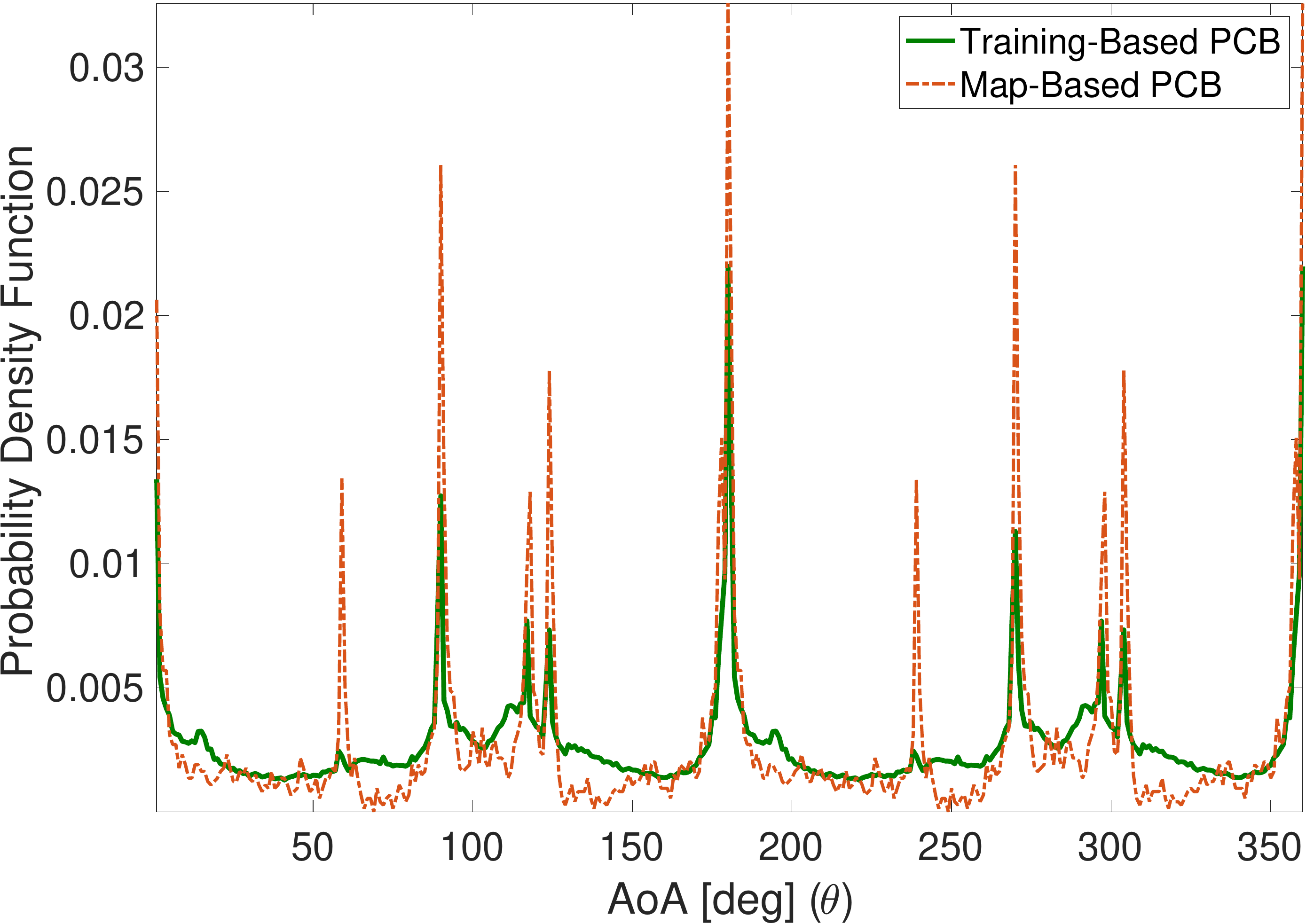}
    \caption{Azimuth probability densities for roundabout.}
    \label{fig:pdfs_comp_round}
\end{figure}

A comparison of the estimated azimuthal angle Probability Density Functions (PDFs) of the two approaches for the intersection scenario in Fig.~\ref{fig: intersection} is reported in Fig.~\ref{fig:pdfs_comparison}. As it can be seen, the Map-based PDF follows the trend of the one obtained through simulations. While for the roundabout scenario in Fig.~\ref{fig:roundabout} the Map-PCB present two peaks around $60^\circ$ and $140^\circ$, see  Fig.~\ref{fig:pdfs_comp_round}. This is due to the presence of roads (lined with buildings) with such inclination that affect the estimated AoA/AoD PDFs.
The HT-based PCBs delete the training phase and reduce the amount of sensitive shared information (i.e., vehicles position). However, intuitively, HT can be applied only in case of scenarios with a high density of buildings along the roads, and its PCB lacks in considering the vehicles mobility characteristics.

\subsection{Non-Uniform Quantization Codebook}
By inspecting the distribution of the azimuth AoA/AoD  in Fig.~\ref{fig:pdfs_comparison} and Fig.~\ref{fig:pdfs_comp_round}, it comes that they are
far from being uniform. Therefore, the codebook designed in~\eqref{eq:uniformCB} can be modified to account for the specific angular distribution. 

Here, the Lloyd-Max algorithm in Algorithm~\ref{alg:LLoydMax} is proposed to derive the optimally quantized angles. The Lloyd-Max quantizer is an iterative method that minimizes the mean squared error of the quantized angles according the angular distribution~\cite{lloyd1,lloyd2}. The key idea behind is to give a quantization step that is related to the PCBs, such that a finer granularity is reserved for most probable angles.

\begin{algorithm}[!t] 
\begin{small}
\caption{LLoyd-Max Quantizer}
\label{alg:LLoydMax}
\begin{algorithmic}[1]
\Require{Angular distribution $p_{\vartheta}(\vartheta)$}
\Ensure{Optimal PCB $\boldsymbol{{\vartheta}}$}
\State Initialize $\boldsymbol{\hat{\vartheta}}$ as in \eqref{eq:uniformCB}
\State \textbf{While} $\left\{\sum_{j=0}^{K_c-1} \left(\hat{\vartheta}_j-\vartheta\right)^2 \; p_{\vartheta}(\vartheta) > \epsilon \right\}$
\State \hspace{5mm}\textbf{For} \{$1 \leq j \leq K_c$\}
\State \hspace{10mm}$\hat{t}_j = 0.5 \left(\hat{\vartheta}_{j-1} + \hat{\vartheta}_{j}\right)$
\State \hspace{10mm}$\hat{\vartheta}_{j} = \frac{\sum_{\hat{t}_j}^{\hat{t}_j+1} \vartheta \; p_{\vartheta}(\vartheta)}{\sum_{\hat{t}_j}^{\hat{t}_j+1} p_{\vartheta}(\vartheta) }$
\State sort $\boldsymbol{{\vartheta}}$ based on $p_{\vartheta}(\vartheta)$
\end{algorithmic}
\end{small}
\end{algorithm}

\section{Performance Evaluation }
\label{sec:performanceevaluation}

In this section, we assess the performance of the different beam selection approches described in Sec.~\ref{sec:BSS} for the three simulated mobility scenario as shown in Fig.~\ref{fig:scenarios}.
The simulation parameters for the SUMO trajectory generation and the V2V settings are in Tab.~\ref{tab:V2Xparameters}.

The maximum Effective Isotropic Radiated Power (EIRP) is set according the current urban limitations, while $P_n$ is assumed $85.5$ dBm from reference sensitivity for power class 2 in FR 2~\cite{TS38101}.

\subsection{IA Latency Analysis}
In this subsection the number of beam sweeping attempts (IA trials) before a correct beam alignment for V2V IA is evaluated for the different methods described in Sec.~\ref{sec:BSS} (both the position-assisted and PCBs-based). A comparison is done with the current 5G NR baseline procedure that is described in Sec.~\ref{sec:IAin5GNR}. 
\begin{table}[]
    \centering
    \begin{tabular}{c c}
         
\begin{tabular}{ | c | c | c |}
	\hline
	\textbf{Parameter}  & \textbf{Urban} & \textbf{Highway} \\ \hline 
	Time step & 100 ms & 100 ms\\  
	Time duration & 600 s & 600 s\\  
	Number of vehicles & 218 & 145 \\  
	Vehicles flow & 1.5 veh/s & 2 veh/s \\  
	Maximum speed & 50 km/h & 130 km/h \\ \hline 
	\end{tabular}
\label{tab:SUMOparameter} &

\begin{tabular}{ | c | c | }
	\hline
	\textbf{Parameter}  & \textbf{Value} \\ \hline 
	Max EIRP & 43 dBm \\ 
	$\sigma^2_{n}$ & -85.5 dBm \\ 
	$f_c$ & 28 GHz \\ 
	Bandwidth $B$ & 400 MHz \\ 
	Antennas height (w.r.t. rooftop) & 0.1 m \\ \hline
	\end{tabular}
    \caption{SUMO and V2V communication parameters.}
    \label{tab:V2Xparameters}
    \end{tabular}
\end{table}

If beams are not perfectly aligned there is a performance degradation in terms of SNR, as observed in Fig.~\ref{fig:snr_gps_ideal}.  
The considered metric is the ECDF of the number of required S-SS blocks attempts before successful beam selection.  

To fasten the IA with the adaptive GPS-assisted schemes, we set a tolerance $\epsilon^k=0.5$. The 5G NR codebook is designed as in \eqref{eq:uniformCB} with $N_s=64$, while the $\boldsymbol{\vartheta}_{PCB}$ are got following the algorithms~\ref{alg:PCB-simulated}-\ref{alg:PCB-HT} with codebook depth $K_c=K_a=64$.
\begin{figure}[!t]
\centering
\subfloat[Urban intersection environment \label{ecdf_inter_urban}]{\includegraphics[width=.33\columnwidth, height=5cm]{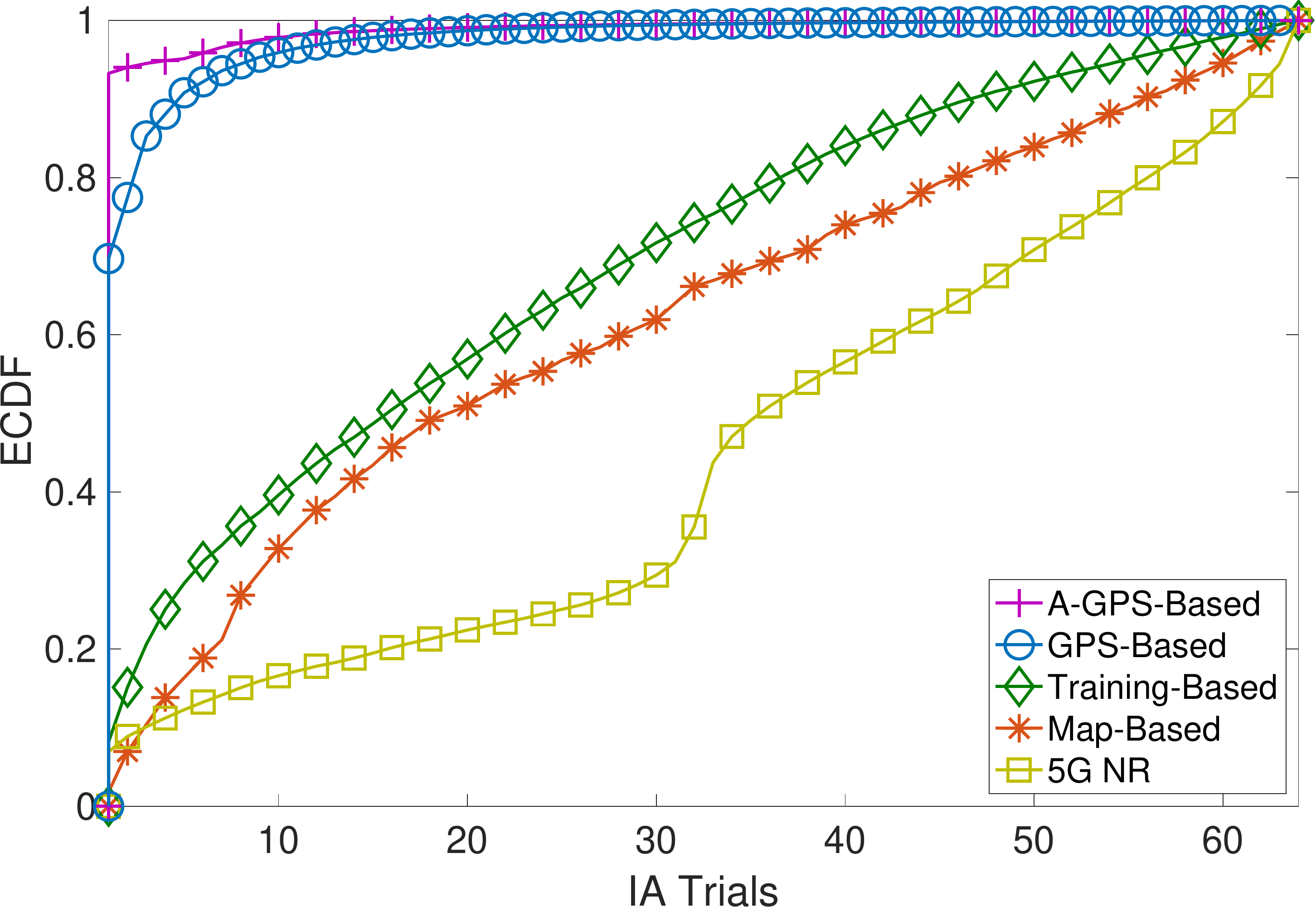}}
\subfloat[Urban roundabout environment \label{ecdf_round_urban}]{\includegraphics[width=.33\columnwidth, height=5cm]{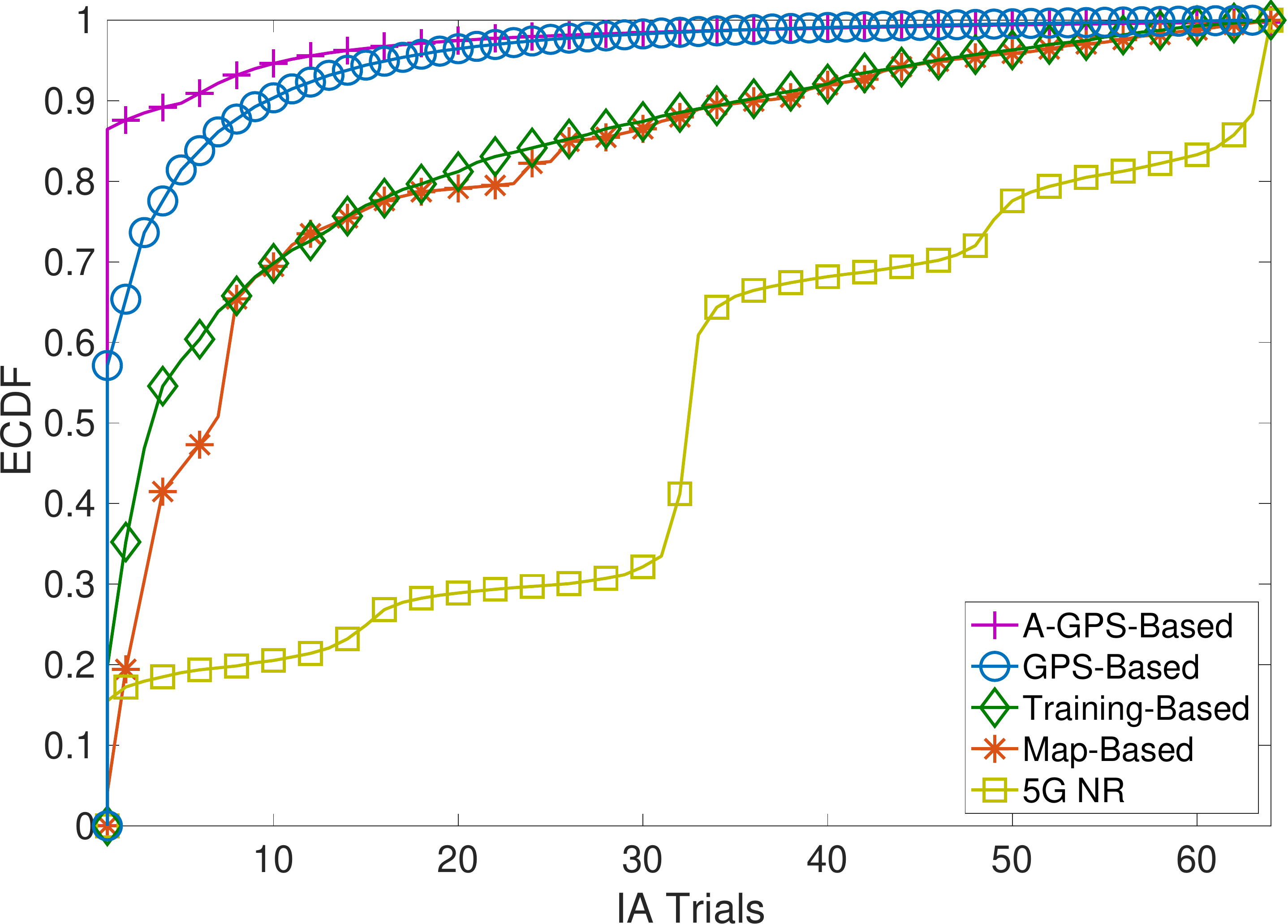}}
\subfloat[Highway environment \label{ecdf_highway}]{\includegraphics[width=.33\columnwidth,height=5cm]{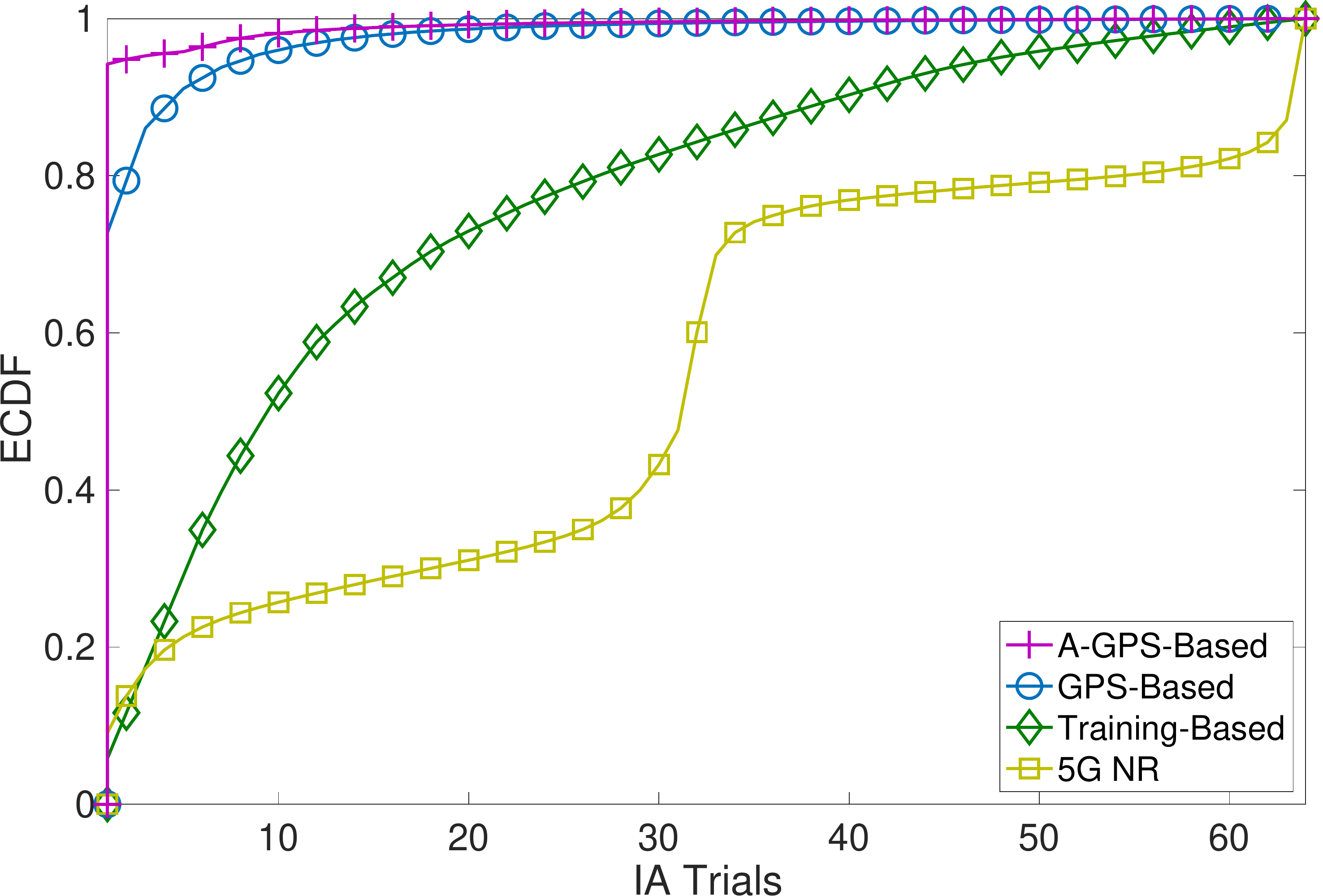}}
\caption{ECDF of the number of required S-SS blocks for a successful IA beam selection for urban intersection (a) and roundabout (b), suburban highway (c) V2V communications. Results are obtained by averaging over 1.6 Millions IA trials.}
\label{ecdf}
\end{figure}

Figure~\ref{ecdf_inter_urban}-\ref{ecdf_round_urban} report the performance for urban scenario, roundabout, and intersection, respectively, while Fig.~\ref{ecdf_highway} shows the highway environment results. 
Generally, in both urban (Fig.~\ref{ecdf_inter_urban}-\ref{ecdf_round_urban}) and highway (Fig.~\ref{ecdf_highway}) scenarios, the current standardized 5G NR procedure shows by far the worst performance, as expected from considerations in Sec. \ref{sec:IAin5GNR}, while position-assisted methods provide faster beam selection. In details, the LMS-GPS adaptive scheme (pink curve), which aims to find the communication direction that maximizes the received power, halves on average the number of trials of the left-right search (blue curve). However, the first requires a computational effort. 
The proposed PCB solutions (simulated and HT-based), instead, even if from one hand show a worse performance with respect to position-assisted schemes, from the other hand they significantly outperforms the 5G NR standard, justifying the intuition behind these scheme, i.e., that in vehicular context beam-based communications can be characterized by prioritized beams, avoiding exhaustive research. 

From the present results it appears that PCBs obtained by simulations outperform the HT-based ones (more in case of roundabout Fig.~\ref{ecdf_round_urban} than intersection Fig.~\ref{ecdf_round_urban}), which as explained in Sec.~\ref{sec:PCB} can be applied only in case of high density building scenarios (not for highway). The slightly gain is obtained by accounting for the vehicles mobility and drivers behavior in the simulated codebook design. The HT-based PCB can not be obtained in case low density of buildings/walls/foliage (e.g., highway), since the angles distribution is got by processing their contours. 

A more detailed analysis of the urban environment  suggests that the PCB (both simulated and HT-based) approaches are more suitable for intersection scenario than for the  roundabout one, since in the latter the probability density of the AoAs/AoDs is more flat (see Fig.~\ref{AoA}). Moreover, in case of intersection the buildings perfectly contour the roads, and therefore, the communication constrains coming from the map topology can be determined using the buildings outline, which is the main focus of HT-based method.

\subsection{Quantization Impact Analysis}

\begin{figure}[t]
\centering
\subfloat[ \label{fig:snr_loss}]{\includegraphics[width=.45\columnwidth]{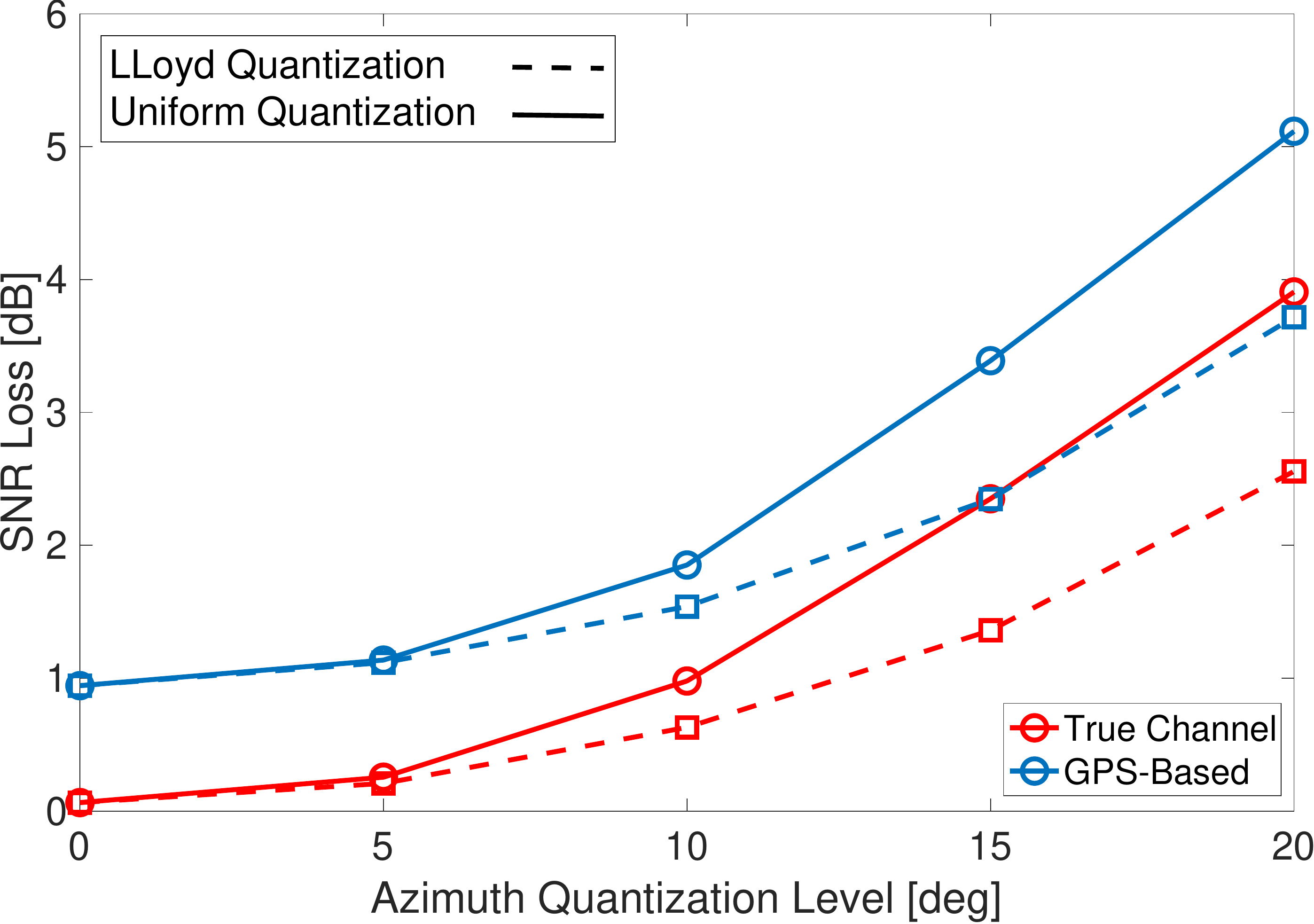}} \hspace{5mm}
\subfloat[ \label{fig_se_loss}]{\includegraphics[width=0.45\columnwidth]{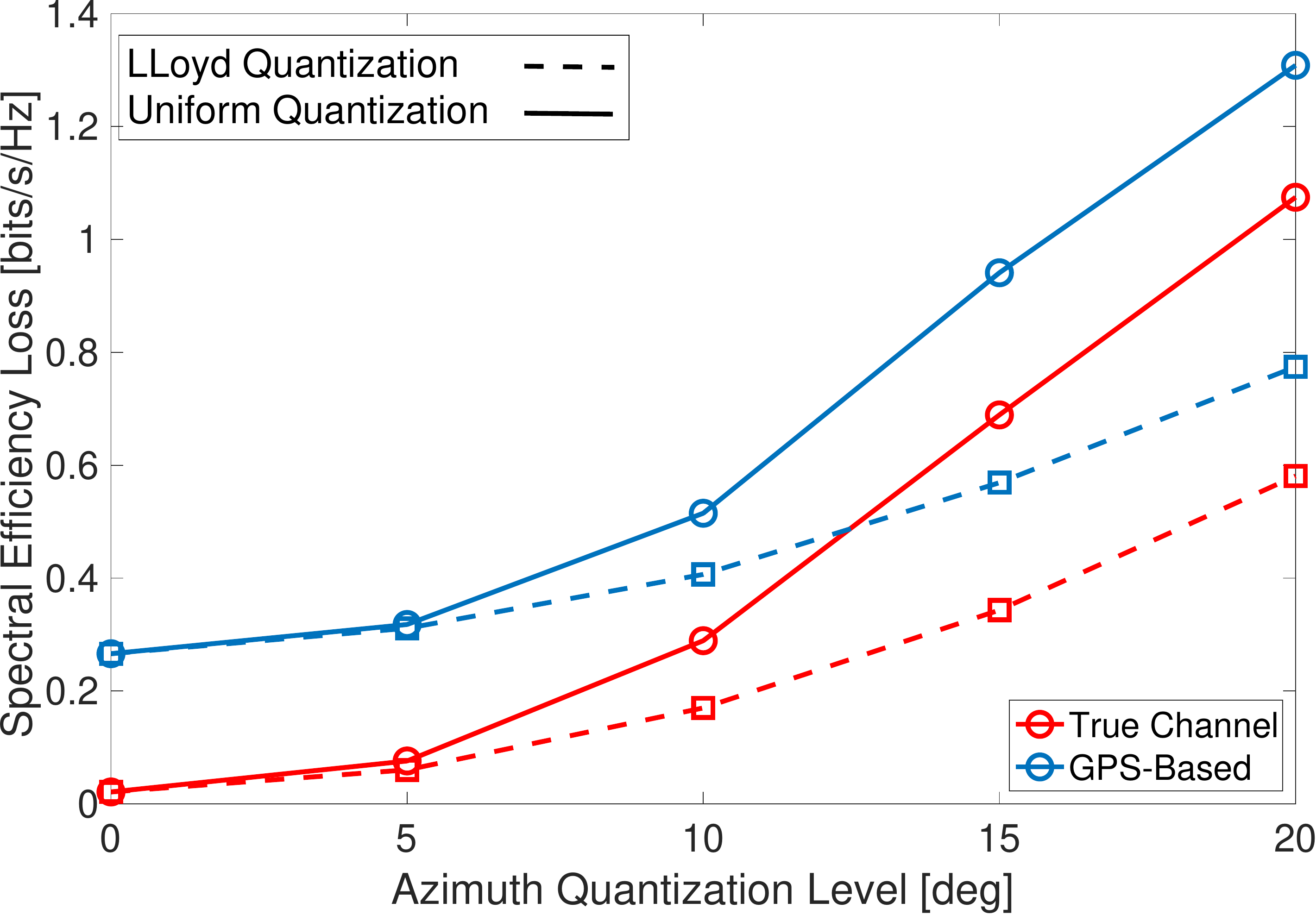}}
\caption{Performance loss after azimuth quantization using uniform and Lloyd’s quantization approaches: (a) SNR loss, (b) SE loss. }
\label{snr_se_loss}
\end{figure}

A performance loss with angle quantization for codebook design is inevitable. As demonstration of the benefit of not-uniform approach presented in Sec.~4.3, the average SNR loss due to quantization is evaluated as 
\begin{align} \label{eq: snr_loss}
  L_{SNR}= \frac{1}{NK} \sum_{i=1}^N \sum_{k=1}^K (\gamma_{opt,i}^k - \gamma_{q,i}^k) \,,
\end{align}
where $\gamma_{opt,i}^k$ is the optimal SNR for $i$th link pair at time slot $k$ derived from SVD method applied to the MIMO channel matrix in \eqref{eq: received signal} and $\gamma_{q,i}^k$ denotes the SNR for $i$th link at time slot $k$ after quantization using uniform or Lloyd’s approaches. 
To compute the optimal SNR in \eqref{eq:SNRdef} we need to know the optimal beamformers. Thus, assuming an ideal knowledge of the channel state information at the transmitter and receiver the SVD of the channel matrix $\mathbf{H}$ is computed as follows
\begin{align} \label{SVD}
    \text{SVD } ({\mathbf{H}})= \mathbf{U}\,\mathbf{D}\,\mathbf{V}^{\text{H}} \,,
\end{align}
where $\mathbf{U}$ and $\mathbf{V}$ are unitary matrices, whose columns are filled with the eigenvector of $\mathbf{H} \mathbf{H}^{\text{H}}$ and $\mathbf{H}^{\text{H}} \mathbf{H}$ respectively, and $\mathbf{D}$ is the diagonal matrix, whose elements are the singular values of $\mathbf{H}$. The optimal beamformers $\mathbf{w}$ and $\mathbf{f}$ coincide with the first column of $\mathbf{U}$ and $\mathbf{V}$, respectively \cite{BrambillaICC2020}.

The average Spectral Efficiency (SE) loss is given by
\begin{align} \label{SE loss}
    L_{SE}= \frac{1}{NK} \sum_{i=1}^N \sum_{k=1}^K (\eta_{opt,i}^k-\eta_{q,i}^k )  \,,
\end{align}
where $\eta_{opt,i}$ is the optimal SE computed using  $\gamma_{opt,i}^k$ for $i$th link pair at time slot k, and $\eta_{q,i}^k$ is the SE for $i$th link at time slot k using uniform or irregular (i.e., Lloyd’s algorithm) quantization. The SE in (\ref{SE loss}) is computed by using the well-known Shannon formula.

To show the benefits of using a non-uniform PCB-based quantization in the simulations, we assume the azimuth angle $\vartheta$ obtained by GPS-information $\mathbf{p}_v(t)$. Thus, we compare performances in case of ideal knowledge of the positions, i.e., $\mathbf{p}_{SUMO}(t)$.
Under the minimum SNR constraint $\gamma_{th}=0$ dB, the SNR loss in (\ref{eq: snr_loss}) and SE loss defined in (\ref{SE loss}) with different quantization levels $\vartheta_q=\{0,5,10,15,20\}$  in degree, where 0 means no quantization implemented) are illustrated in Fig.~\ref{snr_se_loss} (a) and (b), respectively. It can be observed that Lloyd’s algorithm has a smaller and acceptable performance loss compared with uniform quantization in terms of SNR and SE loss. In particular, it can achieve an approximate $1.4$ dB SNR gain and $0.5$ bits/s/Hz SE gain for Lloyd’s algorithm for a 20$^\circ$ quantization level than that using uniform quantization.

\section{Conclusion}
The high data rate and low-latency requirements in the automotive scenarios require  high frequencies sidelink communications. Thus, mmWaves-enabled and MIMO-aided V2V communications will be an integrated part of 6G infrastructure. 
The high mobility of vehicles and roads topology lead to a frequent re-selection of the optimal beam to be used, otherwise a severe link quality degradation is experienced. The beam selection procedure that is performed in the current 5G NR standard, through a periodic and exhaustive search over all possible spatial directions, introduces a significant delay, which can be crucial for advanced driving applications. 
Motivated by this, in this work we presented a number of beam selection techniques that leverage on different performances requirements and shared information.
First, we investigate two position-assisted (i.e., relying on the GPS) cooperative schemes. These approaches can easily satisfy the latency requirements since low trials are required to match the optimal beam starting from the signaled position. However, privacy issues can arise and accurate enough GPS information is not always guaranteed at the physical layer. Then, using the intuition that the constraints coming from road topology result in a non-uniform distribution of the set of  communication directions, we build a Probabilistic Codebook (PCB) approach where the most likely beams are tested first. We propose two different methods to construct PCB. One is based on a training phase, the other uses the Hough Transform to extract the most present straight-lines with their own rotation.

By numerical simulations, we observed PCBs schemes reduce the alignment time more than $80$\% on average with respect to 5G standard in urban (both HT-based and simulated-based) and more than $70$\% on average in suburban scenarios (only simulated-based). 
Finally, we used the PCBs are also used to optimize angles quantization, showing a reduction of the SNR and SE loss with respect to uniform distributed quantization.
As a main contribution we observe that PCBs can be cost-effective for 6G V2V technologies, with the advantage of neither requiring the additional overhead (for knowing the reciprocal position) nor the signaling of privacy-critical information. 
 
\section*{Acknowledgment}
The research has been carried out in the framework of the Huawei-Politecnico di Milano Joint Research Lab. The Authors want to acknowledge Dr. Alberto Perotti, principal research engineer at Huawei and Huawei Milan Research Center.

\bibliographystyle{ieeetr}
\bibliography{Bibliography}

\end{document}

%% file: system/referenceSystem.tex
\begin{figure}[t] \label{fig:angles_ref}
\centering
\begin{tikzpicture}[tdplot_main_coords, scale = 1.5]
\coordinate (P) at ({sqrt(2)/2},{sqrt(2)/2},{0});
\coordinate (o) at (0,0,0);
\coordinate (x) at (1,0,0);
\coordinate (y) at (0,1,0);

\tdplotsetrotatedcoords{0}{0}{0};
\draw[dashed, tdplot_rotated_coords, gray] (0,0,0) circle (1);

\draw[black] (0,0,0) -- (-1.2,0,0);
\draw[black] (0,0,0) -- (0,-1.2,0);

\draw[-stealth] (0,0,0) -- (1.80,0,0) node[below left] {$x$};
\draw[-stealth] (0,0,0) -- (0,1.30,0) node[below right] {$y$};
\draw[-stealth] (0,0,0) -- (0,0,1.30) node[above] {$z$};

\draw (1,1) coordinate (A) -- (0,0) coordinate (B) -- (0,1) coordinate (C) pic ["$\theta$", below right, draw, angle eccentricity=1,fill, blue!60!white] {angle};
\draw (0,1,0) coordinate (A) -- (0,0) coordinate (B) --  ({sqrt(2)/4},{sqrt(2)/4},{sqrt(3)/2}) coordinate (C) pic ["$\phi$", right, draw, angle eccentricity=1,fill, green!70!black] {angle};

\end{tikzpicture}
\caption{Spherical angles reference system.}
\label{fig:reference}
\end{figure}